\documentclass[aps,showpacs,twocolumn,superscriptaddress]{revtex4}

\newcommand{\bea}{\begin{eqnarray}}
\newcommand{\eea}{\end{eqnarray}}
\newcommand{\beq}{\begin{equation}}
\newcommand{\eeq}{\end{equation}}

\usepackage{amsmath}
\usepackage{amssymb}
\usepackage{latexsym}
\usepackage{graphicx}

\usepackage{graphics}
\usepackage{psfig}
\usepackage{epsfig}
\usepackage{color}
\usepackage{changebar}

\begin{document}

%%%%%%%%%%%%%%%%%
%%%   TITLE   %%%
%%%%%%%%%%%%%%%%%

\title{Slice Stretching Effects for Maximal Slicing of a Schwarzschild Black Hole}

%%%%%%%%%%%%%%%%%%%
%%%   AUTHORS   %%%
%%%%%%%%%%%%%%%%%%%

\author{Bernd Reimann}
\affiliation{Max Planck Institut f\"ur Gravitationsphysik,
Albert Einstein Institut, Am M\"uhlenberg 1, 14476 Golm, Germany}
\affiliation{Instituto de Ciencias Nucleares, Universidad Nacional Aut{\'o}noma de M{\'e}xico, A.P. 70-543, M{\'e}xico D.F. 04510, M{\'e}xico}

%%%%%%%%%%%%%%%%
%%%   DATE   %%%
%%%%%%%%%%%%%%%%

\date{October 27, 2004}

%%%%%%%%%%%%%%%%%%%%
%%%   ABSTRACT   %%%
%%%%%%%%%%%%%%%%%%%%

\begin{abstract}
Slice stretching effects such as slice sucking and slice wrapping arise when foliating the extended Schwarzschild spacetime with maximal slices.
For arbitrary spatial coordinates these effects can be quantified in the context of boundary conditions where the lapse arises as a linear combination of odd and even lapse.
Favorable boundary conditions are then derived which make the overall slice stretching occur late in numerical simulations.
Allowing the lapse to become negative, this requirement leads to lapse functions which approach at late times the odd lapse corresponding to the static Schwarzschild metric.
Demanding in addition that a numerically favorable lapse remains non-negative, as result the average of odd and even lapse is obtained. 
At late times the lapse with zero gradient at the puncture arising for the puncture evolution is precisely of this form. 
Finally, analytic arguments are given on how slice stretching effects can be avoided. 
Here the excision technique and the working mechanism of the shift function are studied in detail. 
\end{abstract}

%%%%%%%%%%%%%%%%
%%%   PACS   %%%
%%%%%%%%%%%%%%%%

\pacs{
04.20.Cv,   % fundamental problems and general formulism
04.25.Dm,   % numerical relativity
04.70.Bw,   % classical black holes
95.30.Sf    % relativity and gravitation
\quad Preprint number: AEI-2004-035
}

%%%%%%%%%%%%%%%%%%%%%
%%%   MAKETITLE   %%%
%%%%%%%%%%%%%%%%%%%%%

\maketitle

%%%%%%%%%%%%%%%%%%%%%%%%%%%
\section{Introduction}%%%%%
\label{sec:introduction}%%%
%%%%%%%%%%%%%%%%%%%%%%%%%%%
When a singularity avoiding lapse together with a vanishing shift is used to evolve a spacetime containing a physical singularity, the foliation is of pathological nature \cite{Centrella86,Anninos95,Anninos95_2}.
A first effect referred to as ``slice sucking'' consists of the ``outward''-drifting of coordinate locations as the corresponding Eulerian observers are falling toward the singularity.
This differential infall leads to large proper distances in between neighboring observers, creating large gradients in the radial metric function, the so-called ``slice wrapping''.

The overall effect is referred to as "slice stretching"  and, being a geometric property of the slicing, is present independently of the existence of a numerical grid \cite{commentslicegrid}.
Its appearance has unpleasant consequences when numerically evolving say the Schwarzschild black hole spacetime:
Due to slice sucking the coordinate location of e.g.\ the event horizon is found to move outward and the outer region continuously decreases in coordinate size.
This not only is wasting numerical resources but might cause problems as outer boundary conditions assuming nearly flat space and implemented at a fixed coordinate location become inappropriate and fail.
Furthermore, not being able to resolve with a finite number of grid points the developing steep gradients in components of the 3-metric, numerical inaccuracies caused by slice wrapping force a finite differencing code to crash. 

Following up on earlier work done together with B.~Br\"ugmann, \cite{mythesis,mypaper1,mypaper2}, in the present paper one particular singularity avoiding slicing is looked at, namely maximal slicing corresponding to the condition that the mean extrinsic curvature of the slices vanishes at all times \cite{York79}.
This geometrically motivated choice of the lapse function has been used frequently in numerical relativity, for simulations of a single Schwarzschild black hole see e.g.\ \cite{Estabrook73,Bernstein89,Bernstein93,Bernstein94,Anninos95,Anninos95_2,Daues96,Brewin2001}.

\pagebreak

For Schwarzschild, and by including electric charge also for Reissner-Nordstr\"om, the maximal slices can be constructed analytically \cite{Estabrook73,Reinhart73,Petrich85,Duncan85,Beig98}.
For those spacetimes it is hence possible to examine slice stretching effects on an analytic level.
The discussion throughout this paper is restricted to Schwarzschild, but since the same notation as in \cite{mythesis,mypaper1,mypaper2} is used, the results carry over in a straightforward way to Reissner-Nordstr\"om.

In the following the maximal slices of the Schwarzschild spacetime are re-derived in the radial gauge and the transformation to an Eulerian line element characterized by a vanishing shift is given.
Performing a late time analysis as in \cite{mypaper2} based on \cite{Beig98}, it is shown here that slice stretching arises at the throat of the Einstein-Rosen bridge \cite{Einstein35}.

Assuming symmetry with respect to the throat in order to fix its location in Eulerian coordinates by an isometry condition, it is then possible to quantify slice sucking and wrapping at the event horizon acting as a ``marker''.
Examples of spatial coordinates to be discussed in the context of even boundary conditions are logarithmic grid coordinates (explaining numerical observations of e.g.\ \cite{Bernstein89,Bernstein94}) and isotropic grid coordinates (extending the study of \cite{mypaper2}). 

For boundary conditions other than the even ones, however, this analysis is more involved as one important example, the so-called puncture evolution, shows \cite{mypaper2}.
Here black hole puncture data is evolved using a lapse with zero gradient at the ``puncture'', i.e.\ the compactified left-hand infinity. 
The corresponding lapse function is referred to as "zgp" or puncture lapse.

Focusing on boundary conditions where the lapse arises as superposition of odd and even lapse, two integrals characterizing the overall slice stretching are introduced.

\pagebreak

It is then shown that for ``favorable'' boundary conditions the slice stretching effects can occur arbitrarily late in numerical simulations.
Here the lapse at late times has to approach the odd lapse, the latter corresponding to the static Schwarzschild metric and, being antisymmetric with respect to the throat, having negative values in the left-hand part of the spacetime.
The numerical implementation of a lapse function which is partially negative, however, has been found to be unstable in at least two examples \cite{Brandt94,Koppitz04}.
Demanding hence in addition that a ``numerically favorable'' lapse should be non-negative, for latest possible occurrence of slice stretching a lapse being the average of odd and even lapse is obtained.
One should note here that the puncture lapse is at late times characterized precisely by this condition.
Furthermore, a one-parameter family of boundary conditions ``ranging from odd to even'' is studied numerically and the analytically predicted overall slice stretching is observed.

Finally, analytic arguments are given on how slice stretching effects can be avoided:

By excising parts of the hypersurfaces containing the throat, the latter being the origin of slice stretching, it turns out that analytically the evolution of the metric outside the excision boundary can be expected to freeze in the limit of late times.
Such a behavior has been observed in numerical simulations, see e.g.\ \cite{Seidel92,Anninos95_2}.

Not implementing a ``throat excision'' technique \cite{commentexcision} but making use of the shift function, it turns out that a coordinate singularity in the conformal factor is essential in order to ``hide'' the diverging term of the overall slice stretching there. 
For the puncture evolution this working mechanism is studied in detail for a ``model shift'' freezing the evolution of the rescaled radial metric component and locking the right-hand event horizon. 
For logarithmic grid coordinates, however, the shift function can not be expected to cure slice stretching.

The paper is organized as follows:
In Sec.~\ref{sec:MaximalSlices} the maximal slices of the Schwarzschild spacetime are re-derived and the origin of slice stretching effects is pointed out.
In Sec.~\ref{sec:SliceStretching} slice stretching effects are studied, concentrating in Subsec.~\ref{subsec:evenBCs} on even boundary conditions and discussing logarithmic and isotropic grid coordinates explicitly.
The overall slice stretching is quantified in Subsec.~\ref{subsec:arbitraryBCs} and boundary conditions are derived which make those effects occur late in numerical simulations. 
As examples the puncture lapse and a one-parameter family of boundary conditions are discussed.
In Sec.~\ref{sec:AvoidingSS} analytic arguments are given on how slice stretching effects can be avoided, analyzing in Subsec.~\ref{subsec:excision} the technique of throat excision, and discussing for shift functions in Subsec.~\ref{subsec:logarithmic} their failure and in Subsec.~\ref{subsec:shiftpuncture} their working mechanism for logarithmic and isothermal grid coordinates, respectively.
The results are summarized in Sec.~\ref{sec:conclusion}.

%%%%%%%%%%%%%%%%%%%%%%%%%%%%%%%%%%%%%%%%%%%%%%%%%%%%%%%%%%
\section{Maximal slices of the Schwarzschild spacetime}%%%
\label{sec:MaximalSlices} %%%%%%%%%%%%%%%%%%%%%%%%%%%%%%%%
%%%%%%%%%%%%%%%%%%%%%%%%%%%%%%%%%%%%%%%%%%%%%%%%%%%%%%%%%%
%%%%%%%%%%%%%%%%%%%%%%%%%%%%%
\subsection{Radial gauge}%%%%
\label{subsec:RadialGauge}%%%
%%%%%%%%%%%%%%%%%%%%%%%%%%%%%
Following \cite{Beig98} and starting from the Schwarzschild metric in Schwarzschild coordinates \hbox{$\{t,r,\theta,\phi\}$}, 
\beq
\label{eq:SSmetric}
	ds^{2} = - f(r) dt^{2} + \frac{1}{f(r)} dr^{2} + r^{2}d\Omega^{2} 
        \ \ \ \begin{rm}{with}\end{rm} \ \ \ 
        f(r) = 1 - \frac{2M}{r}, 
\eeq 
the maximal slices of this spacetime are most easily derived as level sets of the form
\beq
\label{eq:levelsets}
	\sigma = t - t(\tau,r) = \begin{rm}{const}\end{rm}.
\eeq
Here the hypersurfaces are labeled by time at infinity $\tau$ and one has to examine the behavior of the normal 
\beq
	n = N \nabla \left( t - t(\tau,r) \right) 
          = N \left( dt - \frac{\partial t}{\partial r} dr \right).
\eeq
Making use of the line element (\ref{eq:SSmetric}), the normalization $N$ is fixed by demanding
\beq
\label{eq:Nnormalization}
	n_{\mu} n^{\mu} = N^{2} \left( -\frac{1}{f(r)} 
                                       + f(r) \left(
 					\frac{\partial t}
                                             {\partial r}\right)^{2} 
					      \right)
                        = -1.
\eeq
As pointed out in \cite{Beig98}, $N$ can also be considered the boost function of the static Killing vector \hbox{$\frac{\partial}{\partial t}$} relative to \hbox{$\sigma = \begin{rm}{const}\end{rm}$},
\beq
\label{eq:NBeig}
	N = -n_{\mu} \left( \frac{\partial}{\partial t} \right)^{\mu}.
\eeq

Calculating the trace of the extrinsic curvature, $K$ turns out to be given by
\beq
\label{eq:Kofnresult}
	K = - \nabla_{\mu} n^{\mu} 
          = \frac{1}{r^{2}} 
	    \frac{d\left[\frac{-r^{2} f(r) \frac{\partial t}{\partial r}}
                              {\sqrt{\frac{1}{f(r)} - f(r)
                               \left( 
			       \frac{\partial t}
                                    {\partial r} \right)^{2} }}
                               \right]}{dr}.
\eeq
Demanding for maximal slicing \hbox{$K \equiv 0$}, the term in the brackets of (\ref{eq:Kofnresult}) obviously has to be a function of time only to be denoted by $C(\tau)$.
Hence  
\beq
\label{eq:dtdr}
	\frac{\partial t}{\partial r}(\tau,r) 
		 = - \frac{C(\tau)}{f(r) \sqrt{r^4 f(r) + C^2(\tau)}} 
\eeq
is found, which has to be integrated by imposing boundary conditions in order to obtain the level sets (\ref{eq:levelsets}).

Furthermore, using the future normal of the foliation, \hbox{$n_{\mu} = - \alpha \nabla_{\mu} \tau$}, and writing the static Killing vector as \hbox{$\left( \frac{\partial}{\partial t} \right)^{\mu} = N n^{\mu} + \xi^{\mu}$} with \hbox{$\xi_{\mu}n^{\mu} = 0$}, the lapse can be obtained as 
\beq
\label{eq:makenewalpha}
	\alpha (\tau,r) = N (\tau,r) \frac{\partial t}{\partial \tau}.
\eeq
Here $N$ by the normalization (\ref{eq:Nnormalization}) is given by
\beq
	N (\tau,r) = \pm \sqrt{f(r) + \frac{C^2(\tau)}{r^4}}
                   = \pm \frac{\sqrt{p_C(r)}}{r^2}
\eeq
when introducing for convenience the polynomial
\beq 
	p_C (r) = r^4 f(r) + C^2 = r^4 - 2M r^3 + C^2, 
\eeq
the subscript $C$ denoting its dependence on $C$ and hence on $\tau$.
Together with the radial metric component
\beq 
\label{eq:gammafinal}
	\gamma (\tau,r) =  \frac{r^4}{p_C(r)}
\eeq
and the shift
\beq
\label{eq:betafinal}
        \beta (\tau,r) = \frac{\alpha(\tau,r)\gamma(\tau,r)}{r^{2}} C(\tau)
\eeq
the maximal slices in the radial gauge
\beq
\label{eq:4mradial}
	ds^{2} = \left( -\alpha^{2} + \frac{\beta^{2}}{\gamma} \right) d\tau^{2} 
                  + 2\beta d\tau dr + \gamma dr^2 + r^{2} d\Omega^{2}
\eeq
have been derived.

Furthermore, as pointed out in \cite{Gentle2001}, the radial and angular components of the extrinsic curvature turn out to be given by
\beq
\label{eq:KAandKB}
	K_r^r = - 2 \frac{C}{r^3}
	\ \ \ \begin{rm}{and}\end{rm} \ \ \ 
	K_\theta^\theta = K_\phi^\phi = \frac{C}{r^3},
\eeq
respectively.

%%%%%%%%%%%%%%%%%%%%%%%%%%%%%%%%%%
\subsection{Eulerian observers}%%%
%%%%%%%%%%%%%%%%%%%%%%%%%%%%%%%%%%
Next the spatial Schwarzschild coordinate $r$ on the maximal slices shall be substituted by a spatial coordinate $z$ corresponding to Eulerian observers.
Applying a transformation of the form \hbox{$r = r(\tau,z)$}, the lapse \hbox{$\alpha = \alpha(\tau,r(\tau,z))$} is still given by (\ref{eq:makenewalpha}) and the line element is characterized by a vanishing shift, 
\beq
\label{eq:4mEuler}
	ds^{2} = -\alpha^{2} d\tau^{2} + G dz^2 + r^{2} d\Omega^{2}.
\eeq
In the context of maximal slicing, \hbox{$K \equiv 0$}, by contracting the evolution equation for the extrinsic curvature, one immediately obtains the statement that for zero shift the determinant of the 3-metric has to be time-independent.
Hence the singularity avoiding property of maximal slicing comes to light as the variation of the local volume remains fixed \cite{Choptuik86}. 
For this reason one can make for the radial metric component the ansatz
\beq
\label{eq:Gis}
	G(\tau,z) = \frac{H(z)}{r^4(\tau,z)}	
\eeq
where the function $H(z)$ depending on $z$ only is determined by the initial data.

The coordinate transformation relating $r$ and $z$ is found by comparison of the radial part of (\ref{eq:4mradial}) and (\ref{eq:4mEuler}).
For fixed slice label $C$ one can infer the ordinary differential equation 
\beq
\label{eq:drdz}
	\left. \frac{d r}{d z} \right| _{C = \begin{rm}{const}\end{rm}} 
            = \pm \frac{\sqrt{p_C(r)}}{r^{4}} \sqrt{H(z)}
\eeq
which can be integrated using the throat as lower integration limit by
\beq
\label{eq:zofr}
        \int\limits_{r_{C}}^{r} \frac{y^{4}\:dy}{\sqrt{p_C(y)}}
	= \pm \int\limits_{z_{C}}^{z} \sqrt{H(y)} \:dy.
\eeq
Here the ``+'' or ``-'' sign applies for the right- or left-hand side of the throat, respectively. 
Furthermore, $r_{C}$ and $z_{C}$ denote the location of the throat in terms of Schwarzschild and Eulerian spatial coordinates. 

Note that $r_{C}$ is found as root of the polynomial \hbox{$p_C(r)$}, which implies \hbox{$C(\tau = 0) = 0$} when starting with the throat of the Einstein-Rosen bridge coinciding initially with the event horizon at \hbox{$r_{EH} = 2M$}.
The throat $r_{C}$, describing the ``innermost'' two-sphere on a slice labeled by $C$, never reaches the singularity at \hbox{$r=0$} as \hbox{$r_{C_{lim}} = 3M/2$} is found in the limit of late times with $C$ approaching 
\beq
\label{eq:Clim}
	C_{lim} = \frac{3}{4}\sqrt{3}M^{2}
\eeq
as pointed out in \cite{Estabrook73}. 
Hence for the Schwarzschild spacetime the singularity avoidance of maximal slices becomes apparent, c.f.\ corollary 3.3 of \cite{Eardley79}.

The coordinate location of the throat in Eulerian coordinates depends in general also on $C$ and is hence a function of time determined by boundary conditions.
Here the behavior of $z_{C}$ can be found by demanding the transformation (\ref{eq:zofr}) to be consistent with the requirement of a vanishing shift as discussed in more detail in \cite{mypaper2}.

By making use of (\ref{eq:Gis}), however, it is possible to describe in the late time limit the profile of the radial metric component near the throat since the latter approaches the value \hbox{$r_{C_{lim}} = 3M/2$} there.

Furthermore, according to (\ref{eq:KAandKB}) for fixed time at infinity the peak in the profiles of the extrinsic curvature components arises at the throat.
For \hbox{$K_\theta^\theta = - K_r^r/2$} in the limit $C \to C_{lim}$ its value there is obtained as \hbox{$C_{lim}/r_{C_{lim}}^3 = 2\sqrt{3}/9M \approx 0.3849/M$}.

%%%%%%%%%%%%%%%%%%%%%%%%%%%%%%%%%%%%%%%%%%%%%%%%%%%%%%%%%
\subsection{Origin and indicators of slice stretching}%%%
\label{subsec:originofss}%%%%%%%%%%%%%%%%%%%%%%%%%%%%%%%%
%%%%%%%%%%%%%%%%%%%%%%%%%%%%%%%%%%%%%%%%%%%%%%%%%%%%%%%%%
As stated in \cite{Beig98}, when demanding antisymmetry with respect to the throat maximal slices are found where $C$ being purely gauge can be chosen independently of time at infinity.
Here the \hbox{3-metric} is given time-independently by the initial data and the odd lapse can be written as 
\beq
\label{eq:oddlapse}
	\alpha_{odd} = \pm \frac{\sqrt{p_C(r)}}{r^2}.
\eeq
In particular, the odd lapse vanishes at the throat and is positive/negative in the original/extended part of the Schwarzschild spacetime to yield the values plus/minus one at right-/left-hand spatial infinity.

Excluding odd boundary conditions where no slice stretching occurs, for a discussion of the late time behavior of maximal slicing instead of the slice label $C$ it turns out to be convenient to introduce $\delta$ as in \cite{Beig98} by
\beq
\label{eq:delta}
	\delta = r_{C} - r_{C_{lim}}. 
\eeq
The late time limit \hbox{$\tau \to \infty$} then can be said to correspond to the limit \hbox{$\delta \to 0$} as the maximal slices approach the limiting slice \hbox{$r = r_{C_{lim}} = 3M/2$} asymptotically. 

By analyzing in this limit the transformation (\ref{eq:zofr}) and the behavior of the 3-metric (\ref{eq:4mEuler}), slice stretching effects can be studied.
As in \cite{mypaper2} this discussion will for simplicity be restricted to the throat and the event horizon acting as markers for slice sucking and wrapping. 

In this reference it has been shown that integrating up to the event horizon, in the limit \hbox{$\delta \to 0$} the integral on the left-hand side of (\ref{eq:zofr}) diverges like 
\beq
\label{eq:originofss}
	  \int\limits_{r_{C}}^{r_{EH}} \frac{y^{4}\:dy}{\sqrt{p_C(y)}} 
        = - C_{lim} \Omega \ln{\left[ \frac{\delta}{M} \right]} + {\cal O}(1)
\eeq
where $\Omega$ is a further fundamental constant given by 
\beq
\label{eq:Omega}
	\Omega = \frac{3}{4}\sqrt{6}M.
\eeq
The proof of this statement is rather lengthy, see \cite{mypaper2} for details.
The divergence proportional to \hbox{$\ln{\left[ \delta \right]}$} can be understood, however, by observing that for \hbox{$C \to C_{lim}$} the lower limit of integration $r_{C}$ becomes a double counting root of the polynomial \hbox{$p_C(r)$}, the root of which appears in the denominator of the integrand of (\ref{eq:originofss}).
Since the upper limit of integration essentially plays no role in this expansion, one should note that any isosurface described by a constant value of \hbox{$r = \begin{rm}{const}\end{rm} \geq r_{EH}$} could be used in the following as a marker for slice stretching effects.
Furthermore, one should note that for the constant (\ref{eq:Omega}) numerically a value of \hbox{$^{0} \Omega \approx 1.82 M$} has been found in \cite{Smarr78b} and that its analytical value has been derived in \cite{Petrich85,Beig98}. 

The diverging term picked up at the throat in (\ref{eq:originofss}) is the origin of slice stretching.
Those effects are hence a feature of the region near the throat.
Denoting the location of the right- and left-hand event horizon by $z^{\pm}_{CEH}$, the subscript $C$ referring again to its time dependence, from the coordinate transformation (\ref{eq:zofr}) in the context of (\ref{eq:originofss}) one can infer 
\beq
\label{eq:zEHofdelta}
	\pm \int\limits_{z_{C}}^{z^{\pm}_{CEH}} \sqrt{H(y)} \:dy
         = - C_{lim} \Omega \ln{\left[ \frac{\delta}{M} \right]} + {\cal O}(1).
\eeq
Hence slice sucking is present as in the limit \hbox{$\delta \to 0$} the event horizon is driven away from the throat by a term diverging logarithmically with $\delta$.

With throat and event horizon moving away from each other, in general also slice wrapping effects in between $z_{C}$ and $z^{\pm}_{CEH}$ show up in the form of an unbounded growth and/or a rapidly steepening gradient in the radial metric component.
To study those, note that in numerical implementations often a time-independent conformal factor $\Psi^{4}(z)$ is factored out from the \hbox{3-metric} to focus on the dynamical features of the metric rather than the static singularity.
In order to discuss the behavior of the rescaled \hbox{3-metric}, it is convenient to introduce 
\beq
\label{eq:definegh}
	g(\tau,z) = \frac{G(\tau,z)}{\Psi^{4}(z)}
	\ \ \ \begin{rm}{and}\end{rm} \ \ \ 
	h(z) = \frac{H(z)}{\Psi^{4}(z)} 
\eeq
which according to (\ref{eq:Gis}) are related by  
\beq
\label{eq:gis}
	g(\tau,z) = \frac{h(z)}{r^{4}(\tau,z)}.	
\eeq

Differentiating now (\ref{eq:gis}) with respect to $z$ by making use of the product rule and (\ref{eq:drdz}), one can extend the study of \cite{mypaper2} by analyzing in addition to the value also the gradient of $g$.
In particular, at the throat one obtains
\beq
\label{eq:dgdzthroat}
	\left. \frac{dg}{dz} \right|_{z_C} 
     =  \frac{1}{r_C^4} \left. \frac{dh}{dz} \right|_{z_C} 
\eeq
since \hbox{$dr/dz$} vanishes there.
Furthermore, at the right- and left-hand event horizon the gradient
\beq
\label{eq:dgdzEH}
	\left. \frac{dg}{dz} \right|_{z^{\pm}_{CEH}} 
      = \frac{1}{r_{EH}^{4}} \left. \frac{dh}{dz} \right|_{z^{\pm}_{CEH}} 
        \mp \frac{4 C h(z^{\pm}_{CEH}) \sqrt{H(z^{\pm}_{CEH})}}
                 {r_{EH}^9} 
\eeq
is found.

Although not obvious from the expressions (\ref{eq:gis}), (\ref{eq:dgdzthroat}) and (\ref{eq:dgdzEH}), one in general can expect both $g$ and \hbox{$dg/dz$} to diverge in the limit of late times at the throat and/or the event horizon.
This happens as the functions $H$, $h$ and \hbox{$dh/dz$} evaluated there usually grow without bounds while the Schwarzschild radius at the throat is approaching $r_{C_{lim}} = 3M/2$ and at the event horizon is given by $r_{EH} = 2M$.
In Sec.~\ref{sec:SliceStretching} slice wrapping will be worked out explicitly for two coordinate choices used frequently in numerical relativity.

In order to describe as a function of time at infinity the slice stretching arising from (\ref{eq:originofss}) and showing up e.g.\ in (\ref{eq:zEHofdelta}), it is in addition necessary to specify the relationship $\delta(\tau)$ by imposing boundary conditions.
Slice stretching for even boundary conditions will be discussed in Subsec.~\ref{subsec:evenBCs} and favorable boundary conditions, where the effects described in terms of $\delta$ show up late in terms of $\tau$, will be derived in Subsec.~\ref{subsec:arbitraryBCs}.

%%%%%%%%%%%%%%%%%%%%%%%%%%%%%%%%%%%%%%%%%%%%%%%%%%%%%%%%%
\section{Slice stretching effects for vanishing shift}%%%
\label{sec:SliceStretching}%%%%%%%%%%%%%%%%%%%%%%%%%%%%%%
%%%%%%%%%%%%%%%%%%%%%%%%%%%%%%%%%%%%%%%%%%%%%%%%%%%%%%%%%
%%%%%%%%%%%%%%%%%%%%%%%%%%%%%%%%%%%%%%%%%%%%%%
\subsection{EVEN BOUNDARY CONDITIONS}%%%%%%%%%
\label{subsec:evenBCs}%%%%%%%%%%%%%%%%%%%%%%%%
%%%%%%%%%%%%%%%%%%%%%%%%%%%%%%%%%%%%%%%%%%%%%%
%%%%%%%%%%%%%%%%%%%%%%%%%%%%%%%%%%%
\subsubsection*{Height function}%%%
%%%%%%%%%%%%%%%%%%%%%%%%%%%%%%%%%%%
Treating the extended and the original part of the Schwarzschild spacetime on equal footing by demanding symmetry with respect to the throat, even boundary conditions are found naturally.
Here in a Carter-Penrose diagram the throat has to remain on the symmetry axis characterized by \hbox{$t = 0$}, see Fig.~4 in \cite{mypaper1}.
As in \cite{Estabrook73,Beig98} by integrating (\ref{eq:dtdr}) for the even ``height function'' one obtains the integral
\beq
\label{eq:teven}    
	t_{even}(C,r) = -\int\limits^{r}_{r_{C}} 
                         \frac{C \:dy}
                              {f(y) \sqrt{p_C(y)}} 
\eeq
defined for \hbox{$r \geq r_{C}$}.
Note that the integration across the pole at $r_{EH}$ is taken in the sense of the principal value and the corresponding slices extend smoothly through both the event horizon $r_{EH}$ and the throat $r_{C}$. 
Since proper time is measured at spatial infinity, from (\ref{eq:teven}) in the limit \hbox{$r \to \infty$} one can infer the relationship
\beq
\label{eq:taueven}
	\tau_{even} (C) = -\int\limits^{\infty}_{r_{C}} 
                           \frac{C \:dy}{f(y) \sqrt{p_C(y)}} 
\eeq
between $\tau_{even}$ and $C$.

%%%%%%%%%%%%%%%%%%%%%%%%%%%%%%%%%%%%%%
\subsubsection*{Late time analysis}%%%
%%%%%%%%%%%%%%%%%%%%%%%%%%%%%%%%%%%%%%
As shown in \cite{Beig98} by expanding $\tau_{even}$ in terms of $\delta$ in the late time limit \hbox{$C \to C_{lim}$}, i.e.\ for \hbox{$\delta \to 0$}, time at infinity is diverging like 
\beq
\label{eq:tauevenofdelta}
	\tau_{even} (\delta)  = -\Omega \ln{\left[ \frac{\delta}{M} \right]} + \Lambda + {\cal O}(\delta)
\eeq
where $\Omega$ has been defined already in (\ref{eq:Omega}) and $\Lambda$ is a constant with the analytic value given by
\beq
\label{eq:Lambda}
	\frac{\Lambda}{M} 
                 = \frac{3}{4}\sqrt{6} \ln{ \left[ 18(3\sqrt{2} - 4) \right]} 
	         - 2 \ln{ \left[ \frac{3\sqrt{3}-5}{9\sqrt{6}-22} \right]}.
\eeq
Here the divergence of $\tau_{even}$ proportional to \hbox{$\ln{\left[\delta\right]}$} arises at the throat for the same reason as pointed out for the expansion (\ref{eq:originofss}).

As later on expansions in $\delta$ will be studied, one should observe that solving in (\ref{eq:tauevenofdelta}) for $\delta$, in leading order with
\beq
\label{eq:deltaeven}
	\frac{\delta}{M} = \exp \left[ \frac{\Lambda}{\Omega} \right]
			   \exp \left[ -\frac{\tau_{even}}{\Omega} \right] 
		          + {\cal O} (\exp \left[ - 2 \frac{\tau_{even}}{\Omega} \right])
\eeq
an exponential decay of $\delta$ with $\tau_{even}$ on the fundamental timescale $\Omega$ is found.

%%%%%%%%%%%%%%%%%%%%%%%%%%%%%%%%%%
\subsubsection*{Lapse function}%%%
%%%%%%%%%%%%%%%%%%%%%%%%%%%%%%%%%%
The even lapse arises from (\ref{eq:makenewalpha}) and is given by
\beq
\label{eq:evenlapse}
	\alpha_{even} = \pm \frac{\sqrt{p_C(r)}}{r^2} \frac{\partial t_{even}}{\partial C}
                          \frac{dC}{d\tau_{even}}.
\eeq
Studying its late time behavior as in \cite{Beig98} and \cite{mypaper2}, it turns out that $\alpha_{even}$ collapses at the throat in order ${\cal O}(\delta)$,
\bea
		\alpha_{even}\mid_{r_{C}} 
	        & = & \frac{2 \sqrt{2}}{3} \frac{\delta}{M} 
			+ {\cal O} (\delta^{2}) \nonumber \\
                & \approx & 0.9428 \frac{\delta}{M} 
			+ {\cal O} (\delta^{2}),  
\eea
and hence in leading order decays exponentially in time on the fundamental timescale $\Omega$ there to avoid the singularity.
By symmetry, at both left- and right-hand event horizon in the limit of late times the finite value \hbox{$C_{lim}/r_{EH}^2 = 3\sqrt{3}/16 \approx 0.3248$} is found, 
\bea
		\alpha_{even}\mid_{r_{EH}} 
		& = & 
                \frac{3}{16}\sqrt{3}
                + \frac{1}{144\sqrt{2}}
		\bigg( 7\sqrt{6} \ln{\left[ 
			\frac{9\sqrt{6}+22}{3\sqrt{2}+4} 
				     \right]} \nonumber \\
                & \ & \ \ + 330 - 36\sqrt{6} - 12\sqrt{3} \bigg)
                \frac{\delta^{2}}{M^{2}} 
			+ {\cal O} (\delta^{3}) \nonumber \\
		& \approx & 0.3248 + 1.2265\frac{\delta^{2}}{M^{2}} 
			+ {\cal O} (\delta^{3}),
\eea
and by construction the even lapse approaches unity at both infinities.
For a particular choice of spatial coordinates, namely isotropic grid coordinates, the time evolution of the even lapse is shown in Fig.~\ref{fig:even}.

%%%%%%%%%%%%%%%%%%%%%%%%%%%%%%%%%%%%%%%%%%%%%%%%%%%%%%%%%%%%%%%
\subsubsection*{Slice stretching in the limit of late times}%%%
%%%%%%%%%%%%%%%%%%%%%%%%%%%%%%%%%%%%%%%%%%%%%%%%%%%%%%%%%%%%%%%
The discussion of slice stretching is particularly simple for even boundary conditions as due to an isometry condition the location of the throat in Eulerian coordinates is given time-independently by its initial value \hbox{$z_{\forall C} = z_{EH} = \begin{rm}{const}\end{rm}$}.
From the expansion (\ref{eq:zEHofdelta}) in the context of (\ref{eq:deltaeven}) one can then infer
\beq
\label{eq:zEHoftaueven}
	\pm \int\limits_{z_{EH}}^{z^{\pm}_{CEH}} \sqrt{H(y)} \:dy
        = C_{lim} \tau_{even} + {\cal O}(1). 
\eeq
Hence in the limit of late times slice sucking is present since the event horizon is driven away from $z_{EH}$ by a term diverging proportional to $\tau_{even}$.

From (\ref{eq:gis}) one can then see that very little evolution is present at the throat since $g$ only grows to \hbox{$(4/3)^4 \approx 3.1605$} times its initial value there as the Schwarzschild radius declines from \hbox{$r_{EH} = 2M$} to \hbox{$r_{C_{lim}} = 3M/2$}.
Furthermore, by a symmetry argument the gradient of $g$ at the throat vanishes.

In order to study slice wrapping at the event horizon, it is then best to look at specific spatial coordinates such as logarithmic or isotropic grid coordinates which both are used frequently in numerical simulations.

%%%%%%%%%%%%%%%%%%%%%%%%%%%%%%%%%%%%%%%%%%%%%%%%%%%%%%%%%%%%%%%
\subsubsection*{First example: Logarithmic grid coordinates}%%%
%%%%%%%%%%%%%%%%%%%%%%%%%%%%%%%%%%%%%%%%%%%%%%%%%%%%%%%%%%%%%%%
Logarithmic grid coordinates $\eta$ \cite{commentgrid} arise when implementing initially the Schwarzschild geometry in terms of logarithmic coordinates corresponding to the \hbox{3-metric} 
\beq
	^{(3)}ds^{2} = \Psi^{4}(\eta) (d\eta^2 + \:d\Omega^{2}).
\eeq
Here at \hbox{$\tau_{even} = 0$} with the conformal factor given by
\beq
\label{eq:Psieta}
	 \Psi(\eta) = \sqrt{2} \cosh{\left[ \frac{\eta}{2} \right]} \sqrt{M}
\eeq
the relationship 
\beq
\label{eq:rofetainitially}
	r(\tau_{even} = 0,\eta) = \Psi^{2}(\eta)
\eeq
between $\eta$ and the Schwarzschild radius $r$ is found.
Independently of time, the throat is located at \hbox{$\eta_{\forall C} = \eta_{EH} = 0$} and the isometry
\beq
\label{eq:etaisometry}
	\eta \longleftrightarrow - \eta
\eeq
is mapping the right-hand part of the spacetime to the left-hand part and vice versa.

With \hbox{$G = \Psi^{12}/r^4$} one may readily verify 
\beq
\label{eq:Feta}
	H(\eta) = \Psi^{12}(\eta)
        \ \ \ \begin{rm}{and}\end{rm} \ \ \ 
        h(\eta) = \Psi^{8}(\eta)
\eeq
and observe that the rescaled radial component of the metric grows from unity to the finite value \hbox{$(4/3)^4 \approx 3.1605$} at the origin being the location of the throat, where in the limit of late times $g$ behaves like \hbox{$\Psi^{8}/r_{C_{lim}}^4$}.
Furthermore, since \hbox{$g = \Psi^{8}/r^4$} is even, obviously its derivative vanishes there.

Discussing slice sucking at the event horizon, from (\ref{eq:zEHoftaueven}) and (\ref{eq:Feta}) it turns out that in leading order the event horizon moves outward like
\beq
\label{eq:etamove}
	\eta^{\pm}_{CEH} \simeq \pm \frac{1}{3} \ln{\left[ \frac{24 C_{lim} \tau_{even}}{M^3} \right]}
                         = {\cal O}\left( \ln{\left[ \tau_{even}^{1/3} \right]} \right). 
\eeq

Inserting this result in (\ref{eq:gis}) while using (\ref{eq:Feta}), slice wrapping is taking place there as $g$ in leading order grows according to
\beq
\label{eq:getablowup}
	\left. g \right|_{\eta^{\pm}_{CEH}}
      = \frac{\Psi^8(\eta^{\pm}_{CEH})}{r_{EH}^4}
      = {\cal O}\left( \tau_{even}^{4/3} \right).
\eeq

Furthermore, making use of (\ref{eq:dgdzEH}) and again (\ref{eq:Feta}), it turns out that a rapidly steepening gradient at the event horizon is present as
\beq
\label{eq:dgdetaEH}
	\left. \frac{dg}{d\eta} \right|_{\eta^{\pm}_{CEH}} 
      \simeq \mp \frac{4 C_{lim} \Psi^{14}(\eta)}{r_{EH}^9}
      = {\cal O}\left( \tau_{even}^{7/3} \right)
\eeq
is found.

These analytical statements should be compared with numerical results as e.g.\ in \cite{Bernstein89,Bernstein94}.
Note that in these simulations it is rather slice wrapping than slice sucking which causes the runs to crash quite early.
This can now be understood by the argument that the event horizon is moving outward only moderately whereas a rapidly steepening gradient in $g$ is found there together with a peak growing slightly further inside.

%%%%%%%%%%%%%%%%%%%%%%%%%%%%%%%%%%%%%%%%%%%%%%%%%%%%%%%%%%%%%%
\subsubsection*{Second example: Isotropic grid coordinates}%%%
%%%%%%%%%%%%%%%%%%%%%%%%%%%%%%%%%%%%%%%%%%%%%%%%%%%%%%%%%%%%%%
Isotropic grid coordinates $x$ \cite{commentgrid} have been constructed in \cite{mypaper1} such that the 4-metric coincides at all times with output from a numerical evolution of black hole puncture data.
Here initially the \hbox{3-metric} 
\beq
	^{(3)}ds^{2} = \Psi^{4}(x) (dx^2 + x^2 \:d\Omega^{2})
\eeq
is implemented making use of isotropic coordinates.
So with the conformal factor 
\beq
\label{eq:Psix}
	\Psi(x) = 1+\frac{M}{2x}
\eeq
it turns out that $x$ at \hbox{$\tau_{even} = 0$} is related to the Schwarzschild radius by 
\beq
\label{eq:rofxinitially}
	r(\tau_{even} = 0,x) = x\Psi^{2}(x).
\eeq
Since for even boundary conditions also during the evolution isotropic and logarithmic grid coordinates are related by
\beq
\label{eq:xeta}
	x = \frac{M}{2} \begin{rm}{e}\end{rm}^{\eta}, 
\eeq
one can see that the region \hbox{$\eta \le 0$} is compactified to \hbox{$0 \le x \le M/2$} and \hbox{$\eta \ge 0$} is mapped to \hbox{$x \ge M/2$}.
Here due to the isometry
\beq
\label{eq:xisometry}
	x \longleftrightarrow \frac{M^2}{4x}
\eeq
the throat is fixed for all times at \hbox{$x_{\forall C} = x_{EH} = M/2$} and the puncture at \hbox{$x = 0$} is simply a compactified image of spatial infinity.

From \hbox{$G = x^4\Psi^{12}/r^4$} then follows
\beq
\label{eq:Fx}
	H(x) = x^4 \Psi^{12}(x)
        \ \ \ \begin{rm}{and}\end{rm} \ \ \ 
	h(x) = x^4 \Psi^{8}(x)
\eeq  
and as for logarithmic grid coordinates one can observe that \hbox{$g = x^4\Psi^8/r^4$} at the throat grows from unity to \hbox{$(4/3)^4 \approx 3.1605$} while due to the isometry a vanishing gradient is present there.

When analyzing slice stretching at the event horizon, it turns out that for the outward-movement of the right-hand event horizon in leading order by using (\ref{eq:zEHoftaueven}) and (\ref{eq:Fx}) (or alternatively (\ref{eq:etamove}) and (\ref{eq:xeta})) for its location 
\beq
\label{eq:xplusmove}
	x^{+}_{CEH} \simeq \left( 3 C_{lim} \tau_{even} \right)^{1/3} 
		    = {\cal O}\left( \tau_{even}^{1/3} \right)
\eeq
is obtained.
The left-hand event horizon, however, due to the isometry (\ref{eq:xisometry}) approaches the puncture like
\beq
\label{eq:xminusmove}
	x^{-}_{CEH} = \frac{M^2}{4 x^{+}_{CEH}}
                    = {\cal O}\left( \tau_{even}^{-1/3} \right).
\eeq

For both right- and left-hand event horizon the rescaled radial metric component $g$ coincides and in leading order diverges according to
\beq
\label{eq:gxblowup}
	\left. g \right|_{x^{\pm}_{CEH}}
	  = \frac{x^{\pm \: 4}_{CEH} \Psi^8(x^{\pm}_{CEH})}{r_{EH}^4}
          = {\cal O}\left( \tau_{even}^{4/3} \right).      
\eeq

Analyzing now the gradient of $g$ at the event horizon, the particular problem of isotropic grid coordinates in the context of even boundary conditions comes to light.
Whereas according to (\ref{eq:dgdzEH}) at the right-hand event horizon the gradient
\beq
	\left. \frac{dg}{dx} \right|_{x^{+}_{CEH}} 
          \simeq - \frac{4 C_{lim} x_{CEH}^{+ \; 6}}{r_{EH}^9}
          = {\cal O}\left( \tau_{even}^2 \right)
\eeq
is found, at the left-hand event horizon the derivative
\bea
	\left. \frac{dg}{dx} \right|_{x^{-}_{CEH}} 
	& = & \frac{dx^{+}_{CEH}}{dx^{-}_{CEH}}
	      \left. \frac{dg}{dx} \right|_{x^{+}_{CEH}} \nonumber \\
        & = & - \frac{M^2}{4x^{- \: 2}_{CEH}} 
	      \left. \frac{dg}{dx} \right|_{x^{+}_{CEH}}
        = {\cal O}\left( \tau_{even}^{8/3} \right)
\eea
is diverging even more rapidly.

\begin{figure}[ht]
	\noindent
	\epsfxsize=80mm \epsfysize=220mm \epsfbox{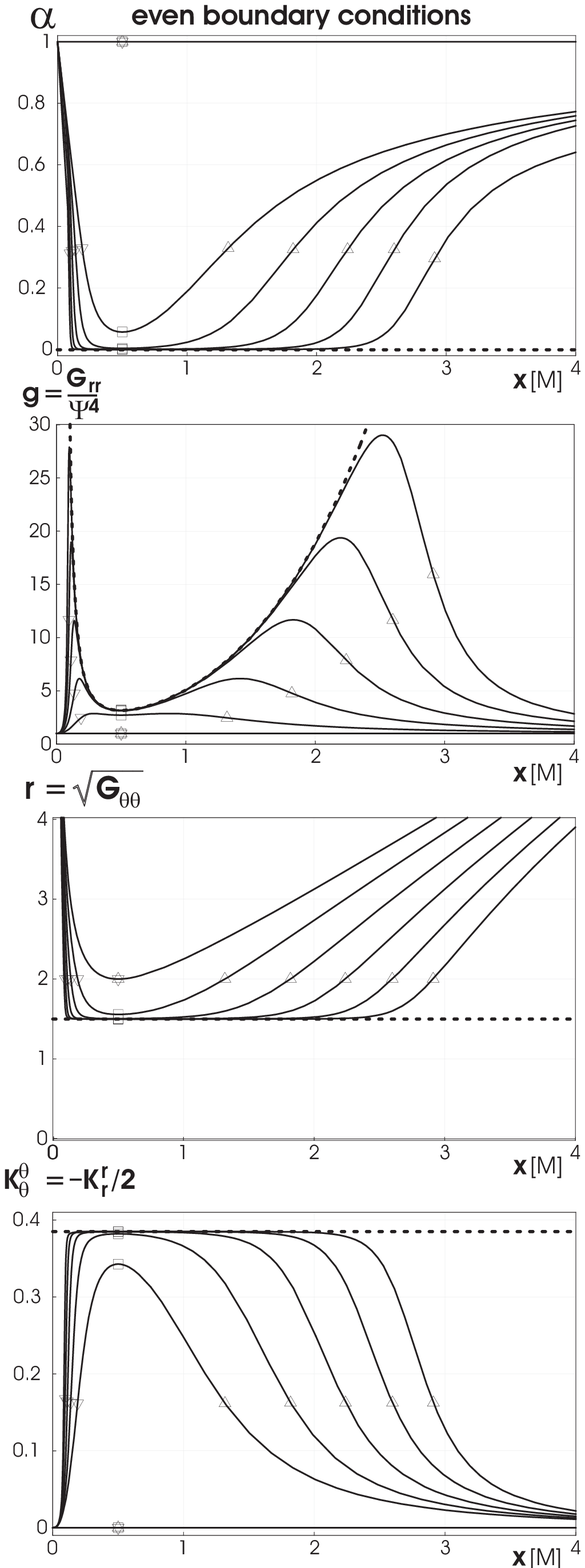}
	\caption{The lapse function and components of both metric and extrinsic curvature are shown as obtained numerically at times \hbox{$\tau_{even} = \{ 0,5M,10M,15M,20M,25M \}$}. The simulation uses even boundary conditions and isotropic grid coordinates corresponding to a vanishing shift function. The location of the throat and the left- or right-hand event horizon is denoted by boxes and down- or upward pointing triangles, respectively. In addition as dotted lines limiting curves are shown which hold in the limit \hbox{$\tau_{even} \to \infty$} near the throat.}
\label{fig:even}	
\end{figure}

As can be seen in Fig.~\ref{fig:even}, a numerically very cumbersome ``double peak'' in the profile of $g$ is developing which due to slice wrapping in the compactified left-hand part of the Schwarzschild spacetime prevents long-lasting simulations.
In Subsec.~\ref{subsec:arbitraryBCs} an analysis of other boundary conditions will show, however, that initially selecting isotropic coordinates can nevertheless be a ``good'' coordinate choice, since the numerically unfavorable behavior described so far can be blamed mainly on the use of even boundary conditions.
Applying for the puncture evolution more adapted ``zgp'' boundary conditions, i.e.\ demanding symmetry at the puncture and obtaining a vanishing gradient of the lapse there, results in significantly better slice stretching behavior as can be seen in Fig.~\ref{fig:zgp}. 

With the Schwarzschild radius at the throat approaching the value \hbox{$r_{C_{lim}} = 3M/2$} in the limit of late times, by making use of \hbox{$g = x^4\Psi^8/r^4$} it is (independently of the boundary conditions) possible to describe the limiting profile of $g$ near the throat by \hbox{$x^4\Psi^8/r_{C_{lim}}^4$}.
The latter has been plotted in both Fig.~\ref{fig:even} and \ref{fig:zgp} as dotted line.   

In addition, extending the study of \cite{mypaper2}, the profiles of the radial and the angular component of the extrinsic curvature shall be discussed here (again for arbitrary boundary conditions).
According to (\ref{eq:KAandKB}), for fixed time at infinity with \hbox{$r \propto \frac{1}{x}$} for \hbox{$\{r \to \infty, x \to 0\}$} both $K_r^r$ and \hbox{$K_\theta^\theta$} near the puncture are of order ${\cal O}(x^3)$, whereas with \hbox{$r = x$} for \hbox{$\{r \to \infty,x \to \infty\}$} for large values of $x$ they decay in order ${\cal O}(x^{-3})$.  
Furthermore, the peak in the profiles of the extrinsic curvature components is found at the throat.
In the late time limit for \hbox{$K_\theta^\theta = - K_r^r/2$} its value there is obtained as \hbox{$C_{lim}/r_{C_{lim}}^3 = 2\sqrt{3}/9M \approx 0.3849/M$}, the latter being in excellent agreement with numerical results as shown in both Fig.~\ref{fig:even} and \ref{fig:zgp}. 

%%%%%%%%%%%%%%%%%%%%%%%%%%%%%%%%%%%%%%%%%%%%%%%%%%%%
\subsection{FAVORABLE BOUNDARY CONDITIONS} %%%
\label{subsec:arbitraryBCs}%%%%%%%%%%%%%%%%%%%%%%%%%
%%%%%%%%%%%%%%%%%%%%%%%%%%%%%%%%%%%%%%%%%%%%%%%%%%%%
%%%%%%%%%%%%%%%%%%%%%%%%%%%%%%%%%%%%%%%%%%%%%%%%%%%%%%%%%%%%%%%%%%%%%%%%%%%%%%
\subsubsection*{Lapse constructed as a superposition of odd and even lapse}%%%
%%%%%%%%%%%%%%%%%%%%%%%%%%%%%%%%%%%%%%%%%%%%%%%%%%%%%%%%%%%%%%%%%%%%%%%%%%%%%%
With the trace of the extrinsic curvature vanishing, \hbox{$K \equiv 0$}, the lapse arises for maximal slicing from the elliptic equation
\beq
\label{eq:dda=Ra} 
	\triangle \alpha = \nabla^{i} \nabla_{i} \alpha = R \alpha
\eeq
where $R$ is the 3-dimensional Ricci scalar.
For fixed time at infinity, this condition is a second order linear ordinary differential equation.
Hence, demanding the lapse to be one at spatial infinity in order to measure proper time there, when supplementing an additional boundary condition the lapse is completely determined.

With odd and even lapse two linearly independent lapse functions satisfying (\ref{eq:dda=Ra}) have been found as pointed out in \cite{Beig98}.
By the superposition principle it is then possible to construct a new lapse, normalized again to unity at right-hand infinity, by a linear combination 
\beq
\label{eq:LK}
	\alpha(\tau,r) = \Phi(\tau) \cdot \alpha_{even}(\tau,r)
                         + (1-\Phi(\tau)) \cdot \alpha_{odd}(\tau,r)
\eeq
with a time-dependent ``multiplicator function'' $\Phi(\tau)$.
An important example for such a superposition is the puncture lapse constructed in \cite{mypaper1}.

%%%%%%%%%%%%%%%%%%%%%%%%%%%%%%%%%%%
\subsubsection*{Height function}%%%
%%%%%%%%%%%%%%%%%%%%%%%%%%%%%%%%%%%
As shown in Subsec.~\ref{subsec:RadialGauge}, the maximal slicing condition fixes the partial derivative of $t$ with respect to $r$ as in (\ref{eq:dtdr}) only, whereas boundary conditions have to be specified to obtain  $t(\tau,r)$ by integration. 
The latter can always be written as the sum of the even height function and a ``time translation function'' $t_{C}(\tau)$ depending on time only,
\beq
\label{eq:generalheightfunction}
	t(\tau,r) =  t_{even}(\tau,r) + t_{C}(\tau),
\eeq
where time at infinity is measured again in the limit \hbox{$r \to \infty$}.
As for \hbox{$\tau = 0$} one starts with the time-symmetric \hbox{$t = 0$} hypersurface, the function $t_{C}$ vanishes initially and is determined during the evolution by boundary conditions.
Furthermore, since the even height function vanishes at the throat during the evolution, $t_{C}$ also represents the value of $t$ at the throat.
Hence the time translation function describes where the throat is found in a Carter-Penrose diagram, see for ``zgp'' boundary conditions Fig.~4 in \cite{mypaper1}.

%%%%%%%%%%%%%%%%%%%%%%%%%%%%%%%%%%%%%%%%%%%%%%
\subsubsection*{Slice stretching integrals}%%%
%%%%%%%%%%%%%%%%%%%%%%%%%%%%%%%%%%%%%%%%%%%%%%
Imposing boundary conditions, the multiplicator function $\Phi(\tau)$ in (\ref{eq:LK}), the time translation function $t_{C}(\tau)$ in (\ref{eq:generalheightfunction}) and the location of the throat in terms of Eulerian coordinates are determined.
Deriving $z_{C}$ as a function of $\tau$ explicitly, however, is rather involved as one has to examine each boundary condition separately when analyzing the coordinate transformation (\ref{eq:zofr}) while making sure that the shift vanishes \cite{mypaper2}.
For a study of slice stretching at e.g.\ the event horizon, though, the location of the throat has to be determined first, since, as shown in Subsec.~\ref{subsec:originofss}, the diverging term in the integral (\ref{eq:zEHofdelta}) being proportional to $\ln{\left[\delta\right]}$ and causing slice stretching is picked up at the throat.

Integrating metric quantities from the left- to the right-hand event horizon yields an alternative approach for a discussion of slice stretching while avoiding in an elegant way inconveniences involved in determining $z_{C}$ as a function of time. 
Although such integrals can not provide items of information like the location of throat or event horizon and value or gradient of the radial metric component there, they are nevertheless excellent indicators for the overall slice stretching.

Here two such integrals shall be introduced, namely
\bea
\label{eq:SH}
	{\cal S}_{H}(\delta) 
	& = & \int\limits_{z^{-}_{CEH}}^{z^{+}_{CEH}} \sqrt{H(y)} \:dy \nonumber \\
        & = & 2 \int\limits_{r_{C}}^{r_{EH}} \frac{y^{4}\:dy}{\sqrt{p_C(y)}} \nonumber \\ 
   	& = & - 2 C_{lim} \Omega \ln{\left[ \frac{\delta}{M} \right]} + {\cal O}(1)
\eea
and
\bea
\label{eq:SG}
	{\cal S}_{G}(\delta) 
	& = & \int\limits_{z^{-}_{CEH}}^{z^{+}_{CEH}} \sqrt{G(\tau,y)} \:dy \nonumber \\
        & = & 2 \int\limits_{r_{C}}^{r_{EH}} \frac{y^{2}\:dy}{\sqrt{p_C(y)}} \nonumber \\ 
	& = &  - 2 \frac{C_{lim} \Omega}{r_{C_{lim}}^2} \ln{\left[ \frac{\delta}{M} \right]} + {\cal O}(1).
\eea
Whereas the first integral arises directly from the coordinate transformation (\ref{eq:zofr}) and essentially has been studied in (\ref{eq:originofss}) and (\ref{eq:zEHofdelta}) already, the second integral is in a more straightforward way related to slice sucking (since the left- and right-hand event horizon appear as limits of integration) and slice wrapping (since the integrand is the root of the radial metric component).
So whereas for some analytical purposes making use of ${\cal S}_{H}$ might be preferable, for numerical studies of slice stretching ${\cal S}_{G}$ should be of particular interest.
Here, in order to locate both left- and right-hand event horizon on the maximal slices to calculate either (\ref{eq:SH}) or (\ref{eq:SG}) numerically, one can compute the Schwarzschild radius $r$ from the prefactor of the angular part of the metric as e.g.\ in \cite{Brandt94,mypaper1} and identify an event horizon as isosurface with \hbox{$r = r_{EH} = 2M$}.
Alternatively, since for the Schwarzschild spacetime event and apparent horizon coincide, for this task both apparent and event horizon finders can be used in principle.

Furthermore, one should note that the shift function does only enter in the calculation of these slice stretching integrals by the fact that for zero shift the function $H$ by construction depends on the spatial coordinate $z$ only, whereas $H$ (like $G$) is a function of both $\tau$ and $z$ otherwise.
For the discussion of Sec.~\ref{sec:AvoidingSS} it is essential to observe here already that the overall slice stretching as defined by (\ref{eq:SH}) and (\ref{eq:SG}) does not depend on the choice of the shift function.

%%%%%%%%%%%%%%%%%%%%%%%%%%%%%%%%%%%%%%%%%%%%%%%%%%%%%%%%%%%%%%%%%%%%%%%%%%%%%%%%
\subsubsection*{Boundary conditions for late observation of slice stretching}%%%
%%%%%%%%%%%%%%%%%%%%%%%%%%%%%%%%%%%%%%%%%%%%%%%%%%%%%%%%%%%%%%%%%%%%%%%%%%%%%%%%
The idea now is to obtain relationships between $\delta$ and $\tau$ which make the overall slice stretching, in terms of $\delta$ arising from (\ref{eq:originofss}) and expressed by integrals such as (\ref{eq:SH}) or (\ref{eq:SG}), occur late in terms of $\tau$ in numerical simulations.
By specifying $\delta(\tau)$, however, boundary conditions for the lapse arise since the multiplicator function appearing in the linear combination (\ref{eq:LK}) can be written as   
\beq
\label{eq:Phi}
        \Phi(\delta) = \frac{\left( \frac{\partial t}{\partial \delta} 
				        - \frac{d\tau}{d\delta}\right)
				   \frac{d\tau_{even}}{d\delta} }				
			   	  {\left( \frac{\partial t_{even}}{\partial \delta} 
				        - \frac{d\tau_{even}}{d\delta}\right)
				   \frac{d\tau}{d\delta}} 
                           = \frac{\frac{d\tau_{even}}{d\delta}}
				  {\frac{d\tau}{d\delta}}. 
\eeq
This expression one can readily verify making use of (\ref{eq:makenewalpha}), (\ref{eq:LK}) and (\ref{eq:generalheightfunction}).
Assuming that a given numerical code for a chosen resolution can only handle a certain amount of overall slice stretching, the hope then is that longer lasting evolutions covering a greater portion of the spacetime can be obtained by imposing more favorable boundary conditions instead of even ones.

As one can see from (\ref{eq:Phi}), demanding that slice stretching effects show up later than for even boundary conditions implies with \hbox{$- d\tau / d\delta > - d\tau_{even} / d\delta > 0$} that the multiplicator function has to be less than one, \hbox{$0 \le \Phi < 1$}.
Note that slice stretching can occur arbitrarily late with $\Phi$ approaching zero and the lapse arising in (\ref{eq:LK}) being then essentially given by the odd lapse.
Furthermore, if $\Phi$ for odd boundary conditions vanishes throughout the evolution, no slice stretching at all is found.

In particular, when looking at \hbox{$\tau = \tau_{even} + t_{C}$}, by making use of (\ref{eq:tauevenofdelta}) one can readily verify that if the time translation function $t_{C}$ is of order ${\cal O}(1)$, as for the even lapse an exponential decay of $\delta$ with $\tau$ is found.
The overall slice stretching then can be expected to be similar to the one arising for even boundary conditions.

Adding, however, a term which diverges logarithmically with $\delta$, i.e.\ \hbox{$t_{C} \simeq -\tilde{\Omega} \ln{\left[ \delta/M \right]}$}, the exponential decay of $\delta$ takes place on a time scale given by the sum of $\Omega$ and $\tilde{\Omega}$. 
The corresponding multiplicator function for this choice of $t_{C}$ is given by \hbox{$\Phi \simeq \Omega/(\Omega + \tilde{\Omega})$}.
In principle with \hbox{$\tilde{\Omega} \to \infty$} the new time scale can be made arbitrarily large to make slice stretching effects occur arbitrarily late.
In this limit, however, one finds that with \hbox{$\Phi \to 0$} the odd lapse is approached.

It is also possible to obtain in leading order not an exponential but a power-law decay of $\delta$ with $\tau$, allowing for very moderate slice stretching behavior. 
Assuming that \hbox{$t_{C} \simeq \delta^{-k}$}, \hbox{$k > 0$}, then \hbox{$\tau \simeq \delta^{-k}$} and \hbox{$\Phi \simeq \Omega \delta^{k} /k$} are found. 
But note that for late times with \hbox{$\delta \to 0$} again the odd lapse is approached.

Whereas it is possible to find boundary conditions such that the overall slice stretching occurs arbitrarily late, the corresponding lapse approaching the odd lapse might from the numerical point of view be disadvantageous as negative values of the lapse occur.
For this reason a numerically favorable lapse in addition should be non-negative.
Using (\ref{eq:LK}) and the formulas (\ref{eq:oddlapse}) and (\ref{eq:evenlapse}) for odd and even lapse, one can show that demanding \hbox{$\alpha \geq 0$} corresponds to \hbox{$\Phi \geq 1/2$}.
Hence the power-law decay of $\delta$ is ruled out, but choosing $0 \le \tilde{\Omega} \le \Omega$ it is possible to find a lapse such that the time scale for the exponential decay of $\delta$ is up to twice the one obtained for even boundary conditions.
In particular, the non-negative lapse showing latest possible occurrence of slice stretching is at late times given by the average of odd and even lapse. 

%%%%%%%%%%%%%%%%%%%%%%%%%%%%%%%%%%%%%%%%%%%%%%%%%%%%%
\subsubsection*{First example: The puncture lapse}%%%
\begin{figure}[ht]
	\noindent
	\epsfxsize=80mm \epsfysize=220mm \epsfbox{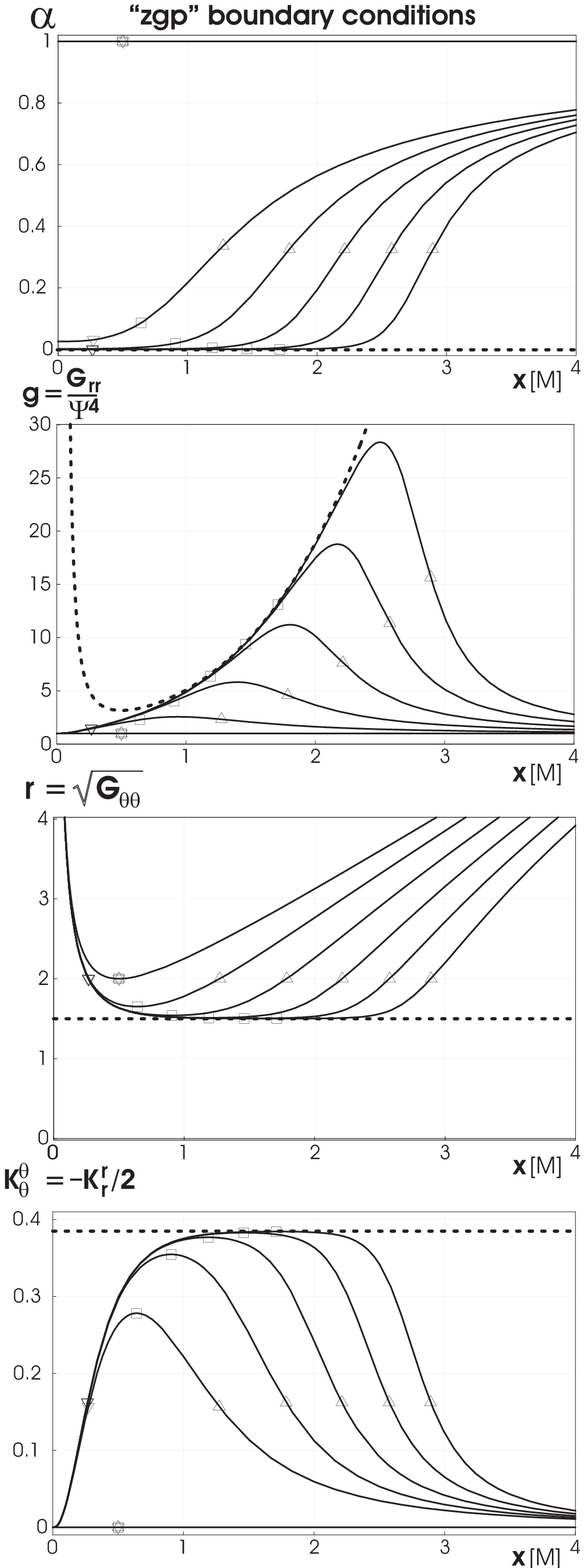}
	\caption{For the puncture evolution corresponding to ``zgp'' boundary conditions, isotropic grid coordinates and zero shift, the geometric quantities as of Fig.~\ref{fig:even} are shown in time steps of \hbox{$\triangle \tau_{zgp} = 5M$}. For \hbox{$\tau_{zgp} \to \infty$} in a region near the throat the puncture lapse collapses to zero, the rescaled radial metric component can be described by $x^4\Psi^8/r_{C_{lim}}^4$, the Schwarzschild radius approaches the value \hbox{$r_{C_{lim}} = 3M/2$} and the angular extrinsic curvature component has the limit \hbox{$C_{lim}/r_{C_{lim}}^3 = 2\sqrt{3}/9M \approx 0.3849/M$}.}
\label{fig:zgp}	
\end{figure}

%%%%%%%%%%%%%%%%%%%%%%%%%%%%%%%%%%%%%%%%%%%%%%%%%%%%%
It is at this point essential to note that the puncture lapse discussed in \cite{mypaper1} is precisely of this form and the puncture evolution of a Schwarzschild black hole is hence taking place in a numerically favorable manner.
As shown in this reference, the multiplicator function of the ``zgp'' lapse is given by
\beq
\label{eq:Phizgp}
	 \Phi = \frac{1}{2}\:\frac{\int\limits^{\infty}_{r_{C}} \frac{y(y-3M)\:dy}
		                  {(y - \frac{3M}{2})^{2} \sqrt{p_C(y)}}}
                                    {\int\limits^{\infty}_{r_{C}} \frac{y(y-3M)\:dy}
		                  {(y - \frac{3M}{2})^{2} \sqrt{p_C(y)}} +\frac{1}{M}}.
\eeq
Here the integral appearing in both the numerator and denominator of (\ref{eq:Phizgp}) is diverging proportional to \hbox{$1/\delta^2$} as the throat in the limit of late times becomes a three-fold root of the denominator of the integrand.
For this reason the ``zgp'' multiplicator function is of the form \hbox{$\Phi_{zgp} = 1/2 + {\cal O}(\delta^{2})$} and the puncture lapse, 
\beq
\label{eq:alphazgp}
  	\alpha^{\pm}_{zgp}(\tau_{zgp},r) 
        = \frac{\sqrt{p_{C}(r)}}{r^{2}}  
	  \frac{\partial t^{\pm}_{zgp}}{\partial C} 
	  \frac{dC}{d\tau_{zgp}}, 
\eeq
being positive everywhere, arises at late times as average of odd and even lapse.
Performing a late time analysis, it turns out that $\alpha_{zgp}$ collapses to zero in order ${\cal O}(\delta^2)$ at the puncture and the left-hand event horizon,
\bea
\label{eq:zgplapsepunctureRN}
		\alpha_{zgp}\mid_{x=0} 
                & = & \frac{2\sqrt{2}}{3}\frac{\delta^{2}}{M^{2}} 
			+ {\cal O} (\delta^{3}) \nonumber \\
		& \approx & 0.9428\frac{\delta^{2}}{M^{2}} 
			+ {\cal O} (\delta^{3}) 
\eea
and
\bea
\label{eq:zgplapseLEHRN}		
		\alpha_{zgp}\mid_{x^{-}_{CEH}} 
		& = &
 		\frac{1}{288\sqrt{2}}
		\bigg(7\sqrt{6} \ln{\left[ 
			\frac{9\sqrt{6}+22}{3\sqrt{2}+4} 
				    \right]} \nonumber \\
		& \ & \ \ + 330 + 60\sqrt{3}\bigg)
                \frac{\delta^{2}}{M^{2}} 
			+ {\cal O} (\delta^{3}) \nonumber \\
		& \approx &
		1.1359\frac{\delta^{2}}{M^{2}} 
			+ {\cal O} (\delta^{3}),
\eea
and in order ${\cal O}(\delta)$ at the throat,
\bea
\label{eq:zgplapsethroatRN}
		\alpha_{zgp}\mid_{x_{C}} 
	        & = & \frac{\sqrt{2}}{3} \frac{\delta}{M} 
			+ {\cal O} (\delta^{2}) \nonumber \\
                & \approx & 0.4714 \frac{\delta}{M} 
			+ {\cal O} (\delta^{2}).  
\eea
At the right-hand event horizon, as for even boundary conditions, the finite value \hbox{$C_{lim}/r_{EH}^2 = 3\sqrt{3}/16 \approx 0.3248$} is found in the limit of late times, 
\bea
\label{eq:zgplapseREHRN}
		\alpha_{zgp}\mid _{x^{+}_{CEH}} 
		& = &
                \frac{3}{16}\sqrt{3}
                + \frac{1}{288\sqrt{2}}
		\bigg( 7\sqrt{6} \ln{\left[ 
			\frac{9\sqrt{6}+22}{3\sqrt{2}+4} 
				     \right]} \nonumber \\
                & \ & \ \ + 330 - 72\sqrt{6} - 12\sqrt{3} \bigg)
                \frac{\delta^{2}}{M^{2}} 
			+ {\cal O} (\delta^{3}) \nonumber \\
		& \approx & 0.3248 + 0.3967\frac{\delta^{2}}{M^{2}} 
			+ {\cal O} (\delta^{3}).
\eea
Finally, the lapse yields unity at infinity in order to measure proper time there, see \cite{mypaper2} for details.
The profile of the puncture lapse for the puncture evolution is shown in Fig.~\ref{fig:zgp}.

In \cite{mypaper1} the ``zgp'' height function is derived as  
\beq
\label{eq:tzgp}
	t_{zgp}^{\pm}(C,r) = t_{even}(C,r) \pm (\tau_{even}(C) - \frac{C}{M})
\eeq
where time is measured at right-hand spatial infinity by
\beq
\label{eq:tauzgp}
	\tau_{zgp}(C) = 2 \tau_{even}(C) - \frac{C}{M}.
\eeq
Making use of the expansion (\ref{eq:tauevenofdelta}) and \hbox{$C = C_{lim} + {\cal O}(\delta^{2})$}, in leading order the exponential decay of $\delta$ with $\tau_{zgp}$ is with
\beq
\label{eq:deltazgp}
	\frac{\delta}{M} = \exp \left[ \frac{2 \Lambda M - C_{lim}}{2 \Omega M} \right]
			   \exp \left[ -\frac{\tau_{zgp}}{2\Omega} \right] 
		          + {\cal O} (\exp \left[ - \frac{\tau_{zgp}}{\Omega} \right])
\eeq
taking place on twice the fundamental timescale $\Omega$.

For the puncture evolution the slice stretching behavior has been discussed in \cite{mypaper2} from left to right at the left-hand event horizon, the throat and the right-hand event horizon.
As shown there, in the limit \hbox{$\tau_{zgp} \to \infty$} the left-hand event horizon is found at a finite value in between its initial location \hbox{$x_{EH} = M/2$} and the puncture, whereas both throat and right-hand event horizon are moving outward like
\beq
\label{eq:xct}
	  x_{C} \simeq \left( \frac{3}{2} C_{lim} \tau_{zgp} \right)^{1/3} 
                = {\cal O} \left( \tau_{zgp}^{1/3} \right)
\eeq
and
\beq
\label{eq:xcplust}
	  x_{CEH}^{+} \simeq \left( 3 C_{lim} \tau_{zgp} \right)^{1/3} 
                      = {\cal O} \left( \tau_{zgp}^{1/3} \right),
\eeq
respectively.

As $x^{-}_{CEH}$ freezes at late times, also the rescaled radial metric component approaches a finite value there.
At $x_{C}$ and $x_{CEH}^{+}$, however, with \hbox{$g = x^4 \Psi^8/r^4$} the metric diverges as in leading order
\beq
\label{eq:gxct}
	\left. g \right|_{x_{C}} 
		\simeq \left( \frac{x_{C}}{r_{C_{lim}}} \right)^4  
	        = {\cal O} \left( \tau_{zgp}^{4/3} \right)
\eeq
and
\beq
\label{eq:gxcplust}
    \left. g \right|_{x_{CEH}^{+}} 
            \simeq \left( \frac{x_{CEH}^{+}}{r_{EH}} \right)^4 
  	    = {\cal O} \left( \tau_{zgp}^{4/3} \right)
\eeq
are found.

Extending the study of \cite{mypaper2}, it turns out that also the gradient of $g$ at the left-hand event horizon freezes, whereas making use of (\ref{eq:dgdzthroat}) and (\ref{eq:dgdzEH}) the derivatives 
\beq
	\left. \frac{dg}{dx} \right|_{x_{C}} 
          \simeq \frac{4 x_{C}^{3}}{r_{C_{lim}}^4}
          = {\cal O}\left( \tau_{zgp} \right)
\eeq
and
\beq
	\left. \frac{dg}{dx} \right|_{x^{+}_{CEH}} 
          \simeq - \frac{4 C_{lim} x_{CEH}^{+ \; 6}}{r_{EH}^9}
          = {\cal O}\left( \tau_{zgp}^2 \right)
\eeq
at the throat and the right-hand event horizon are obtained.

Comparing these ``zgp'' late time statements with the corresponding ones obtained for even boundary conditions in Subsec.~\ref{subsec:evenBCs}, i.e.\ comparing Fig.~\ref{fig:zgp} with Fig.~\ref{fig:even}, one can see that in leading order identical slice sucking and wrapping is present at the right-hand event horizon and to its right.
In the puncture evolution, however, almost no slice stretching occurs to the left of the throat since the ``zgp'' lapse collapses exponentially in time there.
For this reason numerical evolutions of black hole puncture data imposing the ``zgp'' boundary condition are able to last significantly longer than runs forcing even boundary conditions.

%%%%%%%%%%%%%%%%%%%%%%%%%%%%%%%%%%%%%%%%%%%%%%%%%%%%%%%%%%%%%%%%%%%%%%%%%%%%%%%%%%%%%%%%
\subsubsection*{Second example: \hbox{A one-parameter family of boundary conditions}}%%%
\label{subsec:NumRes}%%%%%%%%%%%%%%%%%%%%%%%%%%%%%%%%%%%%%%%%%%%%%%%%%%%%%%%%%%%%%%%%%%%
%%%%%%%%%%%%%%%%%%%%%%%%%%%%%%%%%%%%%%%%%%%%%%%%%%%%%%%%%%%%%%%%%%%%%%%%%%%%%%%%%%%%%%%%
As a further example, a one-parameter family of boundary conditions ranging from odd to even and characterized by a constant multiplicator function in the linear combination (\ref{eq:LK}), \hbox{$\Phi = \begin{rm}{const}\end{rm} \in [0,1]$}, shall be studied numerically in the context of isotropic grid coordinates.
Here the lapse is determined by its time-independent value at the puncture given by 
\beq
\label{eq:alphaatpuncture}
	\alpha(\tau,x=0) = 2 \Phi - 1 \in [-1,1] \ \ \ \forall \tau.
\eeq

The elliptic equation (\ref{eq:dda=Ra}) for the lapse has been implemented in the regularized spherically symmetric code described in \cite{Alcubierre04RegCode}.
Using a shooting method and starting at the puncture with the value (\ref{eq:alphaatpuncture}), to be interpolated there as the origin is staggered in between grid points, the derivative of the lapse has been determined such that when integrating outward a Robin boundary condition \cite{York82} is satisfied.
All simulations shown in this paper have been carried out using $30,000$ grid points for a resolution of \hbox{$\triangle x = 0.001M$} to place the outer boundary at \hbox{$x = 30M$}.

It is worth mentioning that for negative values of the lapse no difficulties have been encountered numerically.
In particular, evolving for odd boundary conditions up to \hbox{$\tau_{odd} = 25M$}, the deviations of lapse and metric components from their initial profiles have been found to be less than $0.1$ per cent.

For the even lapse, however, note that the isometry condition (\ref{eq:xisometry}) has not been enforced actively.
In addition one should then remember that at late times the left-hand event horizon according to (\ref{eq:xminusmove}) gets arbitrarily close to the puncture, whereas the even lapse approaches the value \hbox{$C_{lim}/r_{EH}^2 = 3\sqrt{3}/16 \approx 0.3248$} at $x_{CEH}^{-}$ and is one at \hbox{$x = 0$} as mentioned previously.
Due to the rapidly steepening gradient close to the puncture the shooting method for the even lapse failed shortly after \hbox{$\tau_{even} = 25M$}.

\begin{figure}[ht]
	\noindent
	\epsfxsize=85mm \epsfysize=55mm \epsfbox{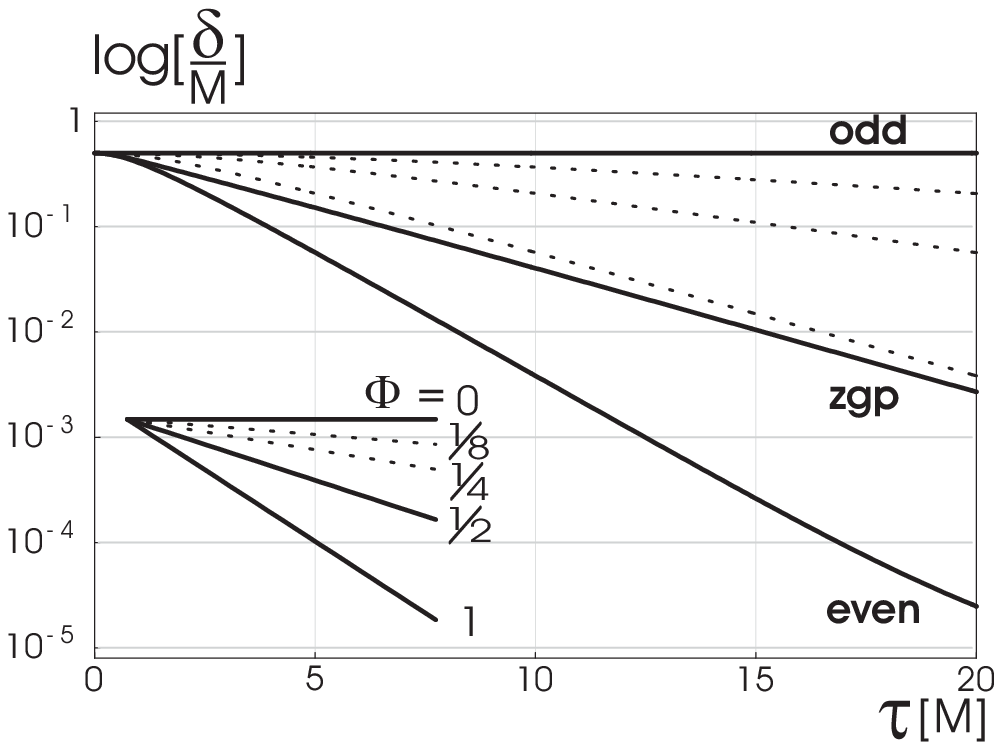}
	\caption{The numerically observed difference in between the Schwarzschild radius at the throat $r_{C}$ and its limiting value \hbox{$r_{C_{lim}} = 3M/2$} is plotted on a logarithmic scale as a function of time. Here the solid lines correspond to odd (\hbox{$\Phi = 0$}), ``zgp'' (\hbox{$\Phi_{zgp} \to 1/2$} in the limit of late times) and even (\hbox{$\Phi = 1$}) boundary conditions. The three dotted curves from top to bottom are characterized by a constant multiplicator function having the value \hbox{$1/8,1/4$} and \hbox{$1/2$}. In the lower left corner the analytically predicted late time slopes are shown.}
\label{fig:ltadelta}	
\end{figure}

In Fig.~\ref{fig:ltadelta} on a logarithmic scale the decay of $\delta$ with time at infinity is shown for runs with constant values \hbox{$\Phi = \left\{ 0,1/8,1/4,1/2,1 \right\}$} and for the puncture evolution where \hbox{$\Phi_{zgp} \to 1/2$} holds in the limit of late times.
In this limit analytically an exponential decay on the timescale \hbox{$\Omega / \Phi$} is predicted, the corresponding slopes are shown in addition in this figure.

\begin{figure}[hb]
	\noindent
	\epsfxsize=85mm \epsfysize=125mm \epsfbox{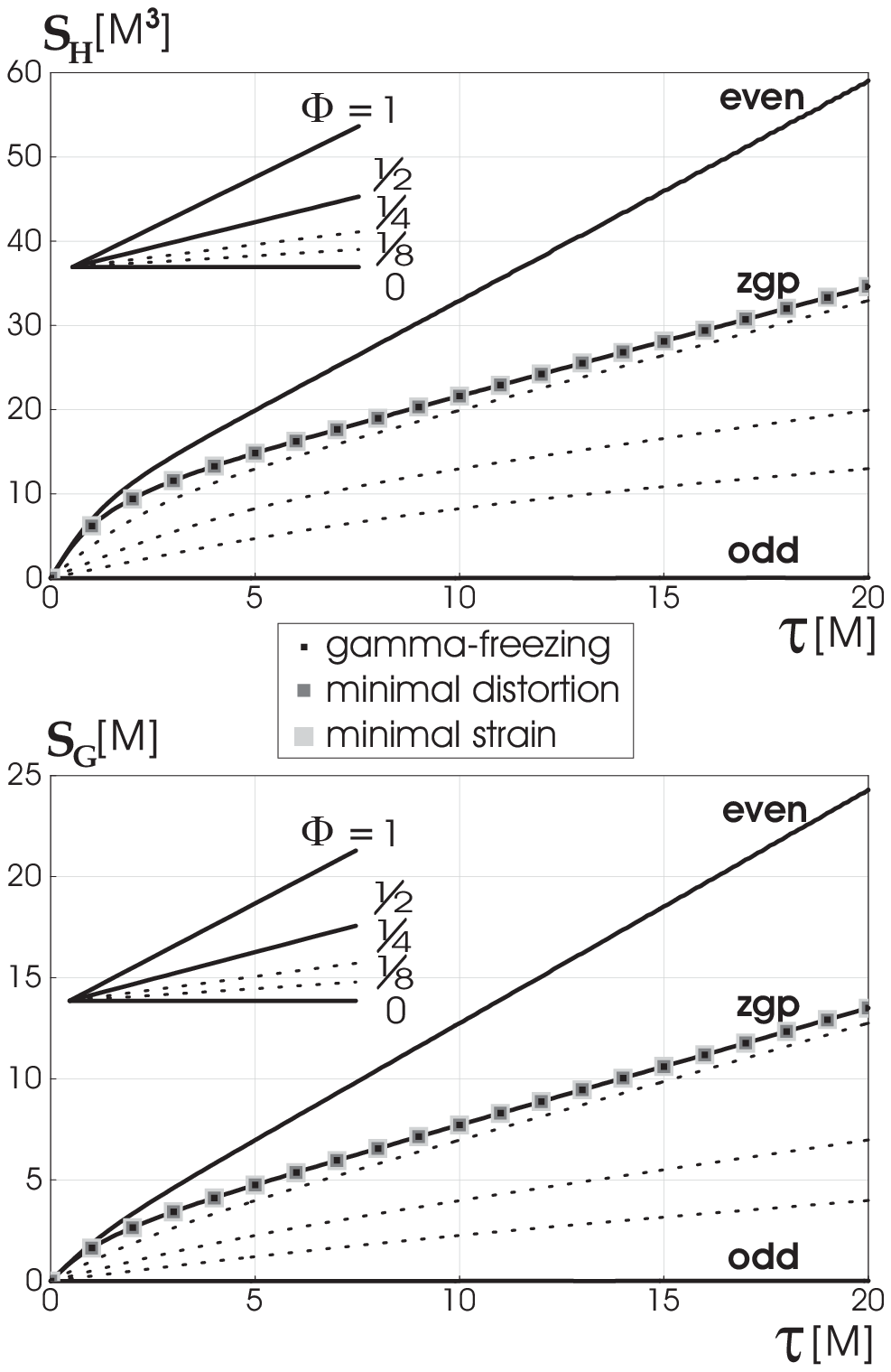}
	\caption{The integrals ${\cal S}_{H}$ and ${\cal S}_{G}$ characterizing the overall slice stretching are shown for the puncture evolution and five members of the one-parameter family of boundary conditions having constant $\Phi$. These curves do not depend on the shift function as shown here explicitly for runs satisfying ``zgp'' boundary conditions and using an elliptic gamma-freezing, minimal distortion or minimal strain shift, see text for details.}
\label{fig:ssintegral}	
\end{figure}

Furthermore, in Fig.~\ref{fig:ssintegral} for the same runs the slice stretching integrals ${\cal S}_{H}$ and ${\cal S}_{G}$, (\ref{eq:SH}) and (\ref{eq:SG}), are plotted together with the expected late time divergence being proportional to time at infinity.

As one can see from these plots, the numerical results are in excellent agreement with analytical predictions.
In particular, one can see that these curves do not depend on the shift function, as runs demanding e.g.\ ``zgp'' boundary conditions and implementing either elliptic gamma-freezing  \cite{Alcubierre02a}, minimal distortion \cite{York79,Bernstein89} or minimal strain \cite{Smarr78b,Bernstein93} shift conditions yield the same results for ${\cal S}_{H}$ and ${\cal S}_{G}$ as simulations using zero shift.

In addition, note that for ``zgp'' boundary conditions as compared to the run with \hbox{$\Phi = 1/2$} more slice stretching arises initially.
This happens since the puncture lapse starts with unit lapse everywhere whereas a ``pre-collapsed'' lapse profile is found when using the average of odd and even lapse from the beginning.
At late times, however, identical slice stretching behavior is found as the two curves become parallel in both Fig.~\ref{fig:ltadelta} and \ref{fig:ssintegral}. 

In the next section analytic arguments will be given on how one can cure slice stretching, analyzing in Subsec.~\ref{subsec:excision} the technique of throat excision, and discussing for shift functions in Subsec.~\ref{subsec:logarithmic} their failure and in Subsec.~\ref{subsec:shiftpuncture} their working mechanism for logarithmic and isothermal grid coordinates, respectively.

%%%%%%%%%%%%%%%%%%%%%%%%%%%%%%%%%%%%%%
\section{Avoiding Slice Stretching}%%%
\label{sec:AvoidingSS}%%%%%%%%%%%%%%%%
%%%%%%%%%%%%%%%%%%%%%%%%%%%%%%%%%%%%%%

%%%%%%%%%%%%%%%%%%%%%%%%%%%%%%%
\subsection{Throat excision}%%%
\label{subsec:excision}%%%%%%%%
%%%%%%%%%%%%%%%%%%%%%%%%%%%%%%%
As studied analytically in Subsec.~\ref{subsec:originofss} and as observed numerically for logarithmic or isotropic/isothermal grid coordinates in \cite{Bernstein89,Bernstein93,Bernstein94,Daues96} or \cite{Anninos95,mythesis,mypaper1,mypaper2}, respectively, slice stretching is arising for maximal slicing in the vicinity of the throat.
Following an idea attributed to W.\ Unruh in \cite{Unruh84} and excising the troublesome region of the hypersurfaces, long-lasting evolutions have been obtained numerically which do not show the previously described slice stretching effects, see e.g.\ \cite{Seidel92,Anninos95_2}.

For an analytic understanding of such excision techniques, consider hypersurfaces starting at some inner excision boundary, given in terms of Schwarzschild and grid coordinates by possibly time-dependent expressions $r_{Cex}$ and $z_{Cex}$, respectively.
Here excision shall take place in between the throat and the right-hand event horizon, so \hbox{$r_C < r_{Cex} < r_{EH}$} and $z_C < z_{Cex} < z^{+}_{CEH}$ hold.
Then one can show that integrals of the form ${\cal S}_{H}$ or ${\cal S}_{G}$, however with the integration starting now at the excision boundary, remain finite as 
\beq
\label{eq:SHexcise}
	\int\limits_{z_{Cex}}^{z^{+}_{CEH}} \sqrt{H(y)} \:dy 
         = \int\limits_{r_{Cex}}^{r_{EH}} \frac{y^{4}\:dy}{\sqrt{p_C(y)}}
   	 = {\cal O}(1)
\eeq
and
\beq
\label{eq:SGexcise}
	\int\limits_{z_{Cex}}^{z^{+}_{CEH}} \sqrt{G(\tau,y)} \:dy 
	 = \int\limits_{r_{Cex}}^{r_{EH}} \frac{y^{2}\:dy}{\sqrt{p_C(y)}} 
	 = {\cal O}(1)
\eeq
are obtained.
Here one should remember that the event horizon as upper integration limit in (\ref{eq:SHexcise}) and (\ref{eq:SGexcise}) - or in (\ref{eq:SH}) and (\ref{eq:SG}) - has been chosen arbitrarily and any other marker lying outside of the excision boundary - or outside of the throat - could be used for such a calculation.
Since integrals like (\ref{eq:SHexcise}) and (\ref{eq:SGexcise}) do not diverge but are finite in the limit of late times, one can now conclude that dynamics present in the 3-metric freeze in this limit.
This has been observed numerically e.g.\ in \cite{Anninos95_2} when implementing horizon-locking coordinates and using several different shift conditions.

In the following no use of throat excision shall be made when discussing in Subsecs.~\ref{subsec:logarithmic} and \ref{subsec:shiftpuncture} the role of a non-trivial shift for evolutions using logarithmic and isothermal grid coordinates, respectively.
 
%%%%%%%%%%%%%%%%%%%%%%%%%%%%%%%%%%%%%%%%%%%%%%%%%%%%%%%%%%%%%%%
\subsection{Shift function and logarithmic grid coordinates}%%%
\label{subsec:logarithmic}%%%%%%%%%%%%%%%%%%%%%%%%%%%%%%%%%%%%%
%%%%%%%%%%%%%%%%%%%%%%%%%%%%%%%%%%%%%%%%%%%%%%%%%%%%%%%%%%%%%%%
For logarithmic grid coordinates $\eta$ the throat is fixed at the origin and, treating both sides of the throat on equal footing by imposing even boundary conditions, the isometry \hbox{$\eta \longleftrightarrow - \eta$} is present during the evolution.

Essentially by making use of the integral ${\cal S}_{H}$, (\ref{eq:SH}), for evolutions with vanishing shift in Subsec.~\ref{subsec:evenBCs} the slice stretching effects have been worked out at the event horizon acting as a marker.
For an arbitrary shift, when integrating the square root of the radial metric component over the throat up to the event horizon, by studying the integral ${\cal S}_{G}$, (\ref{eq:SG}), one can observe that even the use of non-trivial shifts does not change the situation fundamentally: Either slice sucking has to be present as the event horizon is driven away from the origin, or slice wrapping has to occur as the radial metric component blows up, or a combination of both effects is present.

Furthermore, one should note that making use of the time-independent conformal factor $\Psi(\eta)$, (\ref{eq:Psieta}), and evolving \hbox{$g(\tau,\eta) = G(\tau,\eta)/\Psi^4(\eta)$} instead of $G(\tau,\eta)$, does not cure the problem.
This is due to the fact that the conformal factor for logarithmic grid coordinates is finite everywhere, as opposed to the conformal factor (\ref{eq:Psix}) which for puncture data diverges at the puncture, see Subsec.~\ref{subsec:shiftpuncture}.
The particular form of (\ref{eq:Psieta}), however, namely its exponential growth for large values of $\eta$, is responsible for the fact that numerical simulations for logarithmic grid coordinates are supposed to crash due to slice wrapping (leading to large gradients in the radial metric component) rather than due to slice sucking (as the outward-movement of the right-hand event horizon becomes neglectable at late times).

This failure of the shift function to counteract slice stretching when maximally slicing a Schwarzschild black hole in logarithmic grid coordinates has been observed frequently in numerical relativity, but an analytic understanding of this observation was lacking so far.
In particular, the geometrically motivated elliptic minimal distortion shift \cite{Smarr78b,York79} which many people in the numerical relativity community expected to cure the slice stretching problem, numerically had been found to fail.
It is instructive to cite several earlier numerical works here, as these references point out the difficulties encountered numerically when using logarithmic grid coordinates:

D. Bernstein, D. Hobill and L. Smarr reported in \cite{Bernstein89} that ``while the minimal distortion shift vector does reduce a measure of the distortion [...] and removes the sharp gradients of the radial metric component in the region of the event horizon, the coordinates shear is transferred to the throat [...] and produces even larger gradients.'' 
As a consequence, their code when using the minimal distortion shift terminated at about \hbox{$\tau_{even} = 80M$} as compared to runs with zero shift lasting for more than $200M$.
Furthermore, they realized that ``part of the problem may be blamed upon the use of a logarithmic radial variable''.
Similarly, G.\ Daues in \cite{Daues96} found the minimal distortion shift to be ``troublesome in the region near the throat because it leads to a large amount of slice stretching there''. 
Finally, D.\ Bernstein in a numerical study \cite{Bernstein93} of the minimal distortion and various other shift conditions observed that they fail for the Schwarzschild black hole, since ``the minimal distortion gauge is designed to minimize the time derivatives of the conformal metric components only in a volume integral averaged sense''. 

%%%%%%%%%%%%%%%%%%%%%%%%%%%%%%%%%%%%%%%%%%%%%%%%%%%%%%%%%%%%%%
\subsection{Shift function and isothermal grid coordinates}%%%
\label{subsec:shiftpuncture}%%%%%%%%%%%%%%%%%%%%%%%%%%%%%%%%%%
%%%%%%%%%%%%%%%%%%%%%%%%%%%%%%%%%%%%%%%%%%%%%%%%%%%%%%%%%%%%%%
In order to make statements for evolutions taking place in terms of the isothermal grid coordinate $\tilde{x}$, one has to study the transformation of the maximal slices derived in the radial gauge (\ref{eq:4mradial}) to the line element
\beq
\label{eq:4mshift}
      ds^{2} = \left (-A^{2}+\frac{B^{2}}{G} \right) d\tau^{2} 
               + 2B d\tau d\tilde{x} 
               + G d\tilde{x}^2 
               + r^{2} d\Omega^{2}.
\eeq
Here the isothermal grid coordinate $\tilde{x}$ generalizes the isotropic grid coordinate $x$ of previous sections by allowing for a non-vanishing shift $B$.
One can then observe that 
\bea
\label{eq:defineA}
	A(\tau,\tilde{x}) & = &  
		\alpha(\tau,r(\tau,\tilde{x})), \\
\label{eq:defineB}
        B(\tau,\tilde{x}) & = &  
			\beta(\tau,r(\tau,\tilde{x})) 
				\frac{\partial r}{\partial \tilde{x}} 
		      + \gamma(\tau,r(\tau,\tilde{x})) 
				\frac{\partial r}{\partial \tau} 
				\frac{\partial r}{\partial \tilde{x}} 
\eea
and furthermore
\beq
\label{eq:defineG}
	G(\tau,\tilde{x}) 
			 = \Psi^4(\tilde{x}) g(\tau,\tilde{x}) 
			 = \gamma(\tau,r(\tau,\tilde{x}))
			      \left( \frac{\partial r}{\partial \tilde{x}} \right)^{2} 
\eeq
hold.

In the following, the working mechanism of the shift function shall be studied which makes use of the divergence of the conformal factor \hbox{$\Psi(\tilde{x}) = 1 + M/2\tilde{x}$} at the puncture.
For this study the ``zgp'' boundary condition - which in Subsec.~\ref{subsec:arbitraryBCs} has been found to be the numerically favorable boundary condition with latest possible occurrence of slice stretching effects - shall be chosen.
Furthermore, a model shift should act such that during the evolution the rescaled radial metric function freezes with a profile close to one everywhere,
\beq
\label{eq:gone}
	g_{zgp}(\tau_{zgp},\tilde{x}) \approx 1 
            \ \ \forall \tilde{x} 
            \ \ \begin{rm}{as}\end{rm} \ \ \tau_{zgp} \to \infty, 
\eeq
and that with
\beq
\label{eq:REHconst}
	\tilde{x}_{CEH}^{+} (\tau_{zgp}) \approx 
	\begin{rm}{const \ \ as}\end{rm} \ \ \tau_{zgp} \to \infty 
\eeq
the right-hand event horizon at late times is found at a fixed location.
Note that for this analysis the shift has been specified by its desired action rather than by demanding a particular geometric condition as in \cite{mythesis}. 

With these assumptions one can then observe from 
\bea
	\frac{{\cal S}_G}{2} 
	& = &
	\int\limits_{\tilde{x}_{C}(\tau_{zgp})}^{\tilde{x}^{+}_{CEH} 
		\approx \begin{rm}{const}\end{rm}} 
		\Psi^2(y) \sqrt{g_{zgp}(\tau_{zgp},y)} \:dy  \\ \nonumber 
	& = &
	\frac{C_{lim}}{2 r_{C_{lim}}^2} \tau_{zgp} + {\cal O}(1) 	
\eea
and
\bea
	{\cal S}_G 
	& = &
	\int\limits_{\tilde{x}^{-}_{CEH}(\tau_{zgp})}^{\tilde{x}^{+}_{CEH} 
		\approx \begin{rm}{const}\end{rm}} 
		\Psi^2(y) \sqrt{g(\tau_{zgp},y)} \:dy \\ \nonumber 
	& = & \frac{C_{lim}}{r_{C_{lim}}^2} \tau_{zgp} + {\cal O}(1)	
\eea
that both throat and left-hand event horizon are driven toward the puncture since the conformal factor $\Psi(\tilde{x})$ diverges there.
At late times in leading order this movement has to take place according to
\beq
\label{eq:throatmove}
	\tilde{x}_{C}(\tau_{zgp}) \simeq \frac{r_{C_{lim}}^2}{2 C_{lim}} 
				 \frac{M^2}{\tau_{zgp}} 
			  = {\cal O} \left( \tau_{zgp}^{-1} \right) 
\eeq
and
\beq
\label{eq:LEHmove}
	\tilde{x}^{-}_{CEH}(\tau_{zgp}) \simeq \frac{r_{C_{lim}}^2}{4 C_{lim}} 
				       \frac{M^2}{\tau_{zgp}}
			  = {\cal O} \left( \tau_{zgp}^{-1} \right),
\eeq
respectively.

Using these results, one can now discuss the late time behavior of the \hbox{4-metric} at five markers being puncture, left-hand event horizon, throat, right-hand event horizon and spatial infinity.

In particular, making use of the late time statements of Subsec.~\ref{subsec:arbitraryBCs}, it is immediately possible to sketch for the puncture lapse $A_{zgp}$ the corresponding lapse profile.
Here a ``collapsing shoulder'' is obtained which, however, does not move outward as the location of the right-hand event horizon has been locked.

The rescaled radial metric component $g_{zgp}$ by construction behaves nicely everywhere, and the angular metric component is given by the squared value of the Schwarzschild radius $r$.
At the puncture, \hbox{$\tilde{x} = 0$}, and infinitely far to the right, \hbox{$\tilde{x} \to \infty$}, the left- and right-hand spatial infinities are found with $r^2$ behaving like $\tilde{x}^{-2}$ and $\tilde{x}^2$, respectively.   
Furthermore, as both left- and right-hand event horizon correspond to the value \hbox{$r_{EH} = 2M$} and as in between at the throat the limiting value \hbox{$3M/2$} is approached, for the angular metric component the square of those values is obtained at the corresponding locations.

Finally the late time behavior of the shift shall be determined at those five markers.
The calculations, however, are rather lengthy and involve formulas such as (\ref{eq:alphazgp}), (\ref{eq:betafinal}) and (\ref{eq:tzgp}), (\ref{eq:tauzgp}) together with (\ref{eq:teven}), (\ref{eq:taueven}) which have been derived previously for the lapse $\alpha_{zgp}$, the shift $\beta_{zgp}$ and the height function $t_{zgp}$.

According to (\ref{eq:defineB}) and (\ref{eq:defineG}) one can write $B_{zgp}$ as
\beq
\label{eq:Bzgp}        
	B_{zgp} = \Psi^2(\tilde{x}) \sqrt{g_{zgp}} \frac{r^2}{\sqrt{p_C(r)}}
			\left( 
			\alpha_{zgp} \frac{C}{r^2} 
		      + \frac{\partial r}{\partial \tau_{zgp}}
			\right)
\eeq 
and observe that it is natural to rescale the shift by the conformal factor, 
\beq
	B_{zgp}(\tau_{zgp},\tilde{x}) = \Psi^{2}(\tilde{x}) b_{zgp}(\tau_{zgp},\tilde{x}).	
\eeq
Demanding (\ref{eq:gone}), the rescaled shift $b_{zgp}$ is zero at both puncture and infinity.
To obtain statements at the event horizon, one has to insert \hbox{$r_{EH} = 2M$} in (\ref{eq:Bzgp}) and note that the term \hbox{$\partial r/\partial \tau_{zgp}$} vanishes.
The value of $b_{zgp}$ at the left- and right-hand event horizon then turns out to coincide with the value of the lapse there.
The task of calculating the rescaled shift at the throat is more involved.
One might actually worry that $b_{zgp}$ diverges there as $r_C$ is a root of the polynomial $p_C(r)$, the square root of which appears in the denominator of (\ref{eq:Bzgp}).
This, however, does not happen since the term in the brackets of (\ref{eq:Bzgp}) also carries a factor of $\sqrt{p_C(r)}$.
In order to calculate \hbox{$\partial r_C/\partial \tau_{zgp}$}, the chain rule can be applied when differentiating \hbox{$p_{C}(r_C) = 0$} w.r.t.\ $\tau_{zgp}$, obtaining
\beq
	\frac{\partial r_C}{\partial \tau_{zgp}}
	= -\frac{C}{r_C^2 (2r_C - 3M)}
	   \frac{dC}{d\tau_{zgp}}.	
\eeq
Making then in addition use of (\ref{eq:tzgp}) and (\ref{eq:tauzgp}), it finally turns out that the rescaled shift at the throat is given by
\beq
	\left. b_{zgp} \right| _{r_C}
	= \frac{1}{2}
          \frac{\int\limits^{\infty}_{r_{C}} \frac{y(y-3M)\:dy}
		                                  {(y - \frac{3M}{2})^{2} \sqrt{p_C(y)}} + 2/M}
               {\int\limits^{\infty}_{r_{C}} \frac{y(y-3M)\:dy}
		                                  {(y - \frac{3M}{2})^{2} \sqrt{p_C(y)}} + 1/M}
	  \frac{C}{r_C^2}. 
\eeq 
Since here the integral appearing in both the numerator and denominator diverges in the limit of late times, for $b_{zgp}$ at the throat the value \hbox{$C_{lim}/2r_{C_{lim}}^2 = \sqrt{3}/6 = 0.2887$} is found for \hbox{$\tau_{zgp} \to \infty$}. 

There are several comments one should make here regarding these results.
Whereas the angular metric component $r^2$ and the rescaled shift $b_{zgp}$ have been found to behave nicely, the variables $r^2/\Psi^4(\tilde{x})$ and $B_{zgp}$ usually implemented numerically can be expected to develop pathologies close to the puncture. 
Those quantities should blow up at the left-hand event horizon and the throat, which are moving toward the puncture according to (\ref{eq:LEHmove}) and (\ref{eq:throatmove}), in order ${\cal O}\left( \tau_{zgp}^{-4} \right)$ and ${\cal O} \left( \tau_{zgp}^{-2} \right)$, respectively. 
Whereas this behavior might arise for the model shift only, the overall result nevertheless indicates that a shift has to yield ``large values'' close to the puncture as the diverging term of the overall slice stretching is ``hidden'' there.

%%%%%%%%%%%%%%%%%%%%%%%%%%%%%%%%%%%
\section{Conclusion and Outlook}%%%
\label{sec:conclusion}%%%%%%%%%%%%%
%%%%%%%%%%%%%%%%%%%%%%%%%%%%%%%%%%%
Slice stretching effects have been described which show up when maximally slicing the extended Schwarzschild spacetime.
Excluding odd boundary conditions where the static Schwarzschild metric is obtained and no slice stretching occurs, slice sucking and wrapping have been shown to arise at the throat of the maximal slices. 
In terms of $\delta$ in leading order in the limit \hbox{$\delta \to 0$} the overall slice stretching has been characterized by integrals such as ${\cal S}_{H}$ and ${\cal S}_{G}$.
For even boundary conditions and two particular coordinate choices, namely logarithmic and isotropic grid coordinates, slice sucking and wrapping has been worked out explicitly in the context of vanishing shift.  

Searching for favorable boundary conditions, it turned out that slice stretching effects described in terms of $\delta$ can show up arbitrarily late in terms of $\tau$ in numerical simulations if the corresponding lapse approaches the odd lapse and hence becomes negative in the left-hand part of the spacetime.
For numerically favorable boundary conditions demanding in addition the lapse to be non-negative, the latest possible occurrence of slice stretching has been found to take place for a lapse being at late times given by the average of odd and even lapse.
The puncture lapse is precisely of this form and the puncture evolution of a Schwarzschild black hole is hence taking place in a numerically favorable manner.
The late time behavior of the latter has been worked out explicitly here.
In addition, numerical simulations have been performed for the puncture evolution and for a one-parameter family ranging from odd to even, and convincing agreement with analytic results has been found.

Furthermore, analytic arguments have been given on how slice stretching effects can be avoided by implementing e.g.\ excision techniques.
When using a shift, however, it turned out to be essential that the conformal factor has a coordinate singularity in order to hide the diverging term of the overall slice stretching there. 
In particular, the failure of a shift function for logarithmic grid coordinates and its working mechanism for isothermal grid coordinates have been pointed out.

It would be interesting to extend these investigations by detailed studies of geometrically motivated shift conditions such as gamma-freezing and minimal distortion.
Further numerical work could also include algebraic slicings of the ``1 + log'' type \cite{Alcubierre02a}.
The latter are used frequently in numerical simulations and in some regards mimic maximal slicing. 

Finally, as slice stretching by an intuitive argument is often attributed to the singularity avoiding behavior of the slicing, for the Schwarzschild spacetime the maximal slices will be compared in \cite{mypaper4} to geodesic slices.
The latter arise for evolutions with unit lapse and vanishing shift and correspond to freely falling observers which hit the Schwarzschild singularity within finite time.

%%%%%%%%%%%%%%%%%%%%%%%%%%%%
%%%   ACKNOWLEDGEMENTS   %%%
%%%%%%%%%%%%%%%%%%%%%%%%%%%%

\bigskip
\acknowledgments
It is a pleasure for me to thank M.~Alcubierre, B.~Br\"ugmann, J.A.~Gonz{\'a}lez and D.~Pollney. 

%%%%%%%%%%%%%%%%%%%%%%
%%%   REFERENCES   %%%
%%%%%%%%%%%%%%%%%%%%%%

\bibliographystyle{apsrev}

\bibliography{myreferences}

\begin{thebibliography}{33}
\bibitem[{com({\natexlab{a}})}]{commentslicegrid}
\bibinfo{note}{For this reason terms like "grid stretching", "grid sucking" and
  "grid wrapping" appearing in some references are misleading and will not be
  used here.}
\bibitem[{com({\natexlab{b}})}]{commentexcision}
\bibinfo{note}{In numerical relativity this technique is usually refered to as
  "singularity excision". However, since in order to avoid slice stretching one
  has to excise the throat rather than the physical singularity, the latter
  being avoided naturally by maximal slicing, to me the term "throat excision"
  seems to be more appropriate in this context.}
\bibitem[{com({\natexlab{c}})}]{commentgrid}
\bibinfo{note}{Here in order to avoid misunderstandings the spatial coordinates
  obtained during the evolution are refered to as "grid coordinates" to
  distinguish them from the initially implemented "coordinates". Note e.g.\
  that unless odd boundary conditions are used neither
  (\ref{eq:rofetainitially}) nor (\ref{eq:rofxinitially}) hold at later times.}
\end{thebibliography}

%%%%%%%%%%%%%%%
%%%   END   %%%
%%%%%%%%%%%%%%%

\end{document}